\documentstyle[aps,preprint,prc]{revtex}

    \input{psfig.sty}

    \newtheorem{theorem}{Theorem}
    
    \tightenlines

\begin{document}

   \newcommand{\newc}{\newcommand}


    \newc{\ncf}{nucleon correlation function}
    \newc{\qcdsr}{QCD sum rules}
    \newc{\m}{\mbox{ }}



    \newc{\chb}{{\!\not\!C}}
    \newc{\pab}{{\!\not\!P}}
    \newc{\tib}{{\!\not\!T}}


    \newc{\gax}{\gamma_5}
      
                                                                     
  \newc{\vpal}{p_{\alpha}}  \newc{\vqal}{q_{\alpha}}  \newc{\vlal}{l_{\alpha}}
  \newc{\vpbe}{p_{\beta}}   \newc{\vqbe}{q_{\beta}}   \newc{\vlbe}{l_{\beta}}
  \newc{\vpmu}{p_{\mu}}     \newc{\vqmu}{q_{\mu}}     \newc{\vlmu}{l_{\mu}}
  \newc{\vpnu}{p_{\nu}}     \newc{\vqnu}{q_{\nu}}     \newc{\vlnu}{l_{\nu}}
  \newc{\vpxi}{p_{\xi}}     \newc{\vqxi}{q_{\xi}}     \newc{\vlxi}{l_{\xi}}
  \newc{\vpet}{p_{\eta}}    \newc{\vqet}{q_{\eta}}    \newc{\vlet}{l_{\eta}}
  \newc{\vplm}{p_{\lambda}} \newc{\vqlm}{q_{\lambda}} \newc{\vllm}{l_{\lambda}}
  \newc{\vprh}{p_{\rho}}    \newc{\vqrh}{q_{\rho}}    \newc{\vlrh}{l_{\rho}}

  \newc{\gaal}{\gamma_{\alpha}}                                        
  \newc{\gabe}{\gamma_{\beta}}   
  \newc{\gamu}{\gamma_{\mu}}                                                    
  \newc{\ganu}{\gamma_{\nu}}     
  \newc{\gaxi}{\gamma_{\xi}}                                                    
  \newc{\gaet}{\gamma_{\eta}}    
  \newc{\galm}{\gamma_{\lambda}}                                               
  \newc{\garh}{\gamma_{\rho}}    

                
  \newc{\cpal}{p^{\alpha}}  \newc{\cqal}{q^{\alpha}}  \newc{\clal}{l^{\alpha}}
  \newc{\cpbe}{p^{\beta}}   \newc{\cqbe}{q^{\beta}}   \newc{\clbe}{l^{\beta}}
  \newc{\cpmu}{p^{\mu}}     \newc{\cqmu}{q^{\mu}}     \newc{\clmu}{l^{\mu}}
  \newc{\cpnu}{p^{\nu}}     \newc{\cqnu}{q^{\nu}}     \newc{\clnu}{l^{\nu}}
  \newc{\cpxi}{p^{\xi}}     \newc{\cqxi}{q^{\xi}}     \newc{\clxi}{l^{\xi}}
  \newc{\cpet}{p^{\eta}}    \newc{\cqet}{q^{\eta}}    \newc{\clet}{l^{\eta}}
  \newc{\cplm}{p^{\lambda}} \newc{\cqlm}{q^{\lambda}} \newc{\cllm}{l^{\lambda}}
  \newc{\cprh}{p^{\rho}}    \newc{\cqrh}{q^{\rho}}    \newc{\clrh}{l^{\rho}}

  \newc{\haal}{\gamma^{\alpha}}
  \newc{\habe}{\gamma^{\beta}}
  \newc{\hamu}{\gamma^{\mu}}
  \newc{\hanu}{\gamma^{\nu}}
  \newc{\haxi}{\gamma^{\xi}}
  \newc{\haet}{\gamma^{\eta}}
  \newc{\halm}{\gamma^{\lambda}}
  \newc{\harh}{\gamma^{\rho}}

    \newc{\tena}[2]{   \sigma_{{#1}{#2}}            }
    \newc{\tenb}[2]{   \sigma_{{#1}{#2}}       \gax }
    \newc{\tenc}[2]{   {#1} {#2} - {#2} {#1}        }
    \newc{\tend}[2]{ ( {#1} {#2} - {#2} {#1} ) \gax }
    \newc{\tene}[5]{ \epsilon_{ {#1}{#2}{#3}{#4} } {#5}^{#3} \gamma^{#4}      }
    \newc{\tenf}[5]{ \epsilon_{ {#1}{#2}{#3}{#4} } {#5}^{#3} \gamma^{#4} \gax }
    \newc{\teng}[3]{ \mathaccent 94 {#3} ( {#3}_{#1} \gamma_{#2} - {#3}_{#2} \gamma_{#1} ) }       
    \newc{\tenh}[3]{ \mathaccent 94 {#3} ( {#3}_{#1} \gamma_{#2} - {#3}_{#2} \gamma_{#1} ) \gamma_5}



    \newc{\nmass}{M_{n}}                      \newc{\pmass}{M_{p}}
    \newc{\resn}{\lambda_{n}}                 \newc{\resp}{\lambda_{p}}
    \newc{\alphan}{\alpha_{n}}                \newc{\alphap}{\alpha_{p}}

    \newc{\ncon}{s_{0}^{n}}                   \newc{\pcon}{s_{0}^{p}}
    \newc{\betan}{\beta_{n}}                  \newc{\betap}{\beta_{p}}

    \newc{\ncharge}{Q_{n}}                    \newc{\pcharge}{Q_{p}}
    \newc{\namm}{{\mu}_{n}^{a}}               \newc{\pamm}{{\mu}_{p}^{a}}
    \newc{\nedm}{d_{n}}                       \newc{\pedm}{d_{p}}
    \newc{\nmm}{{\mu}_{n}}                    \newc{\pmm}{{\mu}_{p}}
                        
    \newc{\nfc}{F_{1}^{n}(0)}                 \newc{\pfc}{F_{1}^{p}(0)}
    \newc{\nfm}{F_{2}^{n}(0)}                 \newc{\pfm}{F_{2}^{p}(0)}
    \newc{\nfe}{F_{3}^{n}(0)}                 \newc{\pfe}{F_{3}^{p}(0)}
    \newc{\nfa}{F_{4}^{n}(0)}                 \newc{\pfa}{F_{4}^{p}(0)}


    \newc{\massq}{m_q}    \newc{\raq}{R_q}    \newc{\chq}{e_q}
    \newc{\massu}{m_u}    \newc{\rau}{R_u}    \newc{\chu}{e_u}
    \newc{\massd}{m_d}    \newc{\rad}{R_d}    \newc{\chd}{e_d}
    \newc{\masss}{m_s}    \newc{\ras}{R_s}    \newc{\chs}{e_s}
    \newc{\massQ}{M_Q}    \newc{\raQ}{R_Q}    \newc{\chQ}{e_Q}


    \newc{\thetabar} {   { \bar{\theta} }     }
    \newc{\thetabarq}{ { { \bar{\theta} }_q } }
    \newc{\thetabaru}{ { { \bar{\theta} }_u } }
    \newc{\thetabard}{ { { \bar{\theta} }_d } }
    \newc{\thetabars}{ { { \bar{\theta} }_s } }
    \newc{\thetabarQ}{ { { \bar{\theta} }_Q } }

    \newc{\thetag} { { \theta_G }    } 
    \newc{\thetagq}{ { \theta_{Gq} } }
    \newc{\thetagu}{ { \theta_{Gu} } }
    \newc{\thetagd}{ { \theta_{Gd} } }
    \newc{\thetags}{ { \theta_{Gs} } }
    \newc{\thetagQ}{ { \theta_{GQ} } }

    \newc{\thetaq}{ {\theta _ q} }
    \newc{\thetau}{ {\theta _ u} }
    \newc{\thetad}{ {\theta _ d} }
    \newc{\thetas}{ {\theta _ s} }
    \newc{\thetaQ}{ {\theta _ Q} }

    \newc{\cond}[1]{ {\langle {#1} \rangle} _ {\thetaq,\thetag} }
    \newc{\varcond}[3]{ {\langle {#1} \rangle} _ {#2,#3} }


    \newc{\pinu}{\Pi^{N}(p)}
    \newc{\pizero}{\Pi_{0}^{N}(p)}
    \newc{\pimunu}{\Pi_{\mu\nu}^{N}(p)}
   
    \newc{\npi}{\Pi^{n}(p)}
    \newc{\npizero}{\Pi_{0}^{n}(p)}
    \newc{\npimunu}{\Pi_{\mu\nu}^{n}(p)}

    \newc{\ppi}{\Pi^{p}(p)}
    \newc{\ppizero}{\Pi_{0}^{p}(p)}
    \newc{\ppimunu}{\Pi_{\mu\nu}^{p}(p)}
 
    \newc{\deb}{\!\not\!D}
    \newc{\ftr}[2]{     \int d^{2\omega} {#2} \m e^{ i {#1} {#2} }     }

    \draft

    \preprint{\vbox{\it  \null\hfill\rm DOE/ER/41014-13-N97}\\\\}

    \title{ Nucleon Electric Dipole Moments from QCD Sum Rules }

    \author{ Chuan-Tsung Chan$^{(1)}$ 
             \thanks{E-mail address: ctchan@phys.ntu.edu.tw}
             \and 
             Ernest M. Henley$^{(2)(3)}$
             \thanks{E-mail address: henley@phys.washington.edu}
             \and
             Thomas Meissner$^{(4)}$
             \thanks{E-mail address: meissner@yukawa.phys.cmu.edu}   }

    \address{ $^{(1)}$Department of Physics,
                      National Taiwan University,
                      Taipei, 10617 Taiwan 
              \and \\
              $^{(2)}$Department of Physics, Box 351560,
                      University of Washington,
                      Seattle, WA 98195, USA
              \and \\
              $^{(3)}$Institute for Nuclear Theory, Box 351550,
                      University of Washington,
                      Seattle, WA 98195, USA 
              \and \\
              $^{(4)}$Department of Physics,
                      Carnegie-Mellon University,
                      Pittsburgh, PA 15213, USA }

    \date{June.30, 1997}

    \maketitle

    \begin{abstract}


The electric dipole moments of nucleons ( NEDM, $d_N$ ) are calculated using 
the method of QCD sum rules. 
Our calculations are based on the parity ( $\pab$ ) and  time reversal 
( $\tib$ ) violating parameter $\thetabar$ in QCD and establish a functional 
dependence of the NEDM on $\thetabar$, without assuming a perturbative 
expansion of this symmetry breaking parameter. 
The results obtained from the QCD sum rules approach are shown to be consistent
with the general symmetry constraints on CP violations in QCD, including the 
necessity of: (1) finite quark masses, (2) spontaneous chiral symmetry 
breaking, and (3) the $U_A(1)$ anomaly.
Given the current experimental upper bound on the neutron electric dipole 
moment ( nEDM ), $ | d_n | \leq 10^{-25} e \cdot cm $, we find 
$ | \thetabar | \leq 10^{-9} $. 
This result is compatible with previous calculations of nEDM using different 
techniques and excludes the possibility of solving the strong CP problem within
QCD via a dynamical suppression mechanism.

    \end{abstract}


    \narrowtext

    \section{Motivation}
    \label{sec:y1}


  In this paper, we study the electric dipole moments of nucleons ( NEDM,
  denoted as $d_N$ ), which serves as an indicator of both parity ( P ) and 
  time reversal ( T ) symmetry breaking \cite{NEDM:ramsey}.
  The main focus is on the possible violations of P and T ( or CP ) symmetries
  in the strong interactions, with Quantum Chromodynamics ( QCD ) as the
  underlying theory.
  In this picture, the nucleons ( N ) are treated as composite particles
  consisting of quarks (~q ) and gluons ( G ) and the strong P and T violating
  interaction in the QCD Lagrangian is characterized by a $\thetabar$ parameter
  \cite{QCD:theta} ( see Sec.\ref{sec:y2} for a definition of this parameter ).
  Such a P and T violating interaction among quarks and gluons generates a
  coupling of the nucleon spin to the external electric field, and the strength
  of this coupling is defined as the electric dipole moment of the nucleon.

  Our purpose is to establish a functional dependence of the NEDM on the 
  $\thetabar$ parameter, along with other fundamental parameters of the QCD
  Lagrangian, e.g., the quark masses $m_q$, and the values of quark condensates
  $R_q$.
  Based on the current experimental upper bound of the neutron electric dipole
  moment ( nEDM, denoted as $d_n$ ), which is less than $10^{-25} e \cdot cm$
  \cite{NEDM:exp}, we can obtain an upper bound on the strong CP 
  violating\footnote{Here we use the CPT theorem to translate time reversal
  noninvariance as CP violation.} parameter $\thetabar$. 
  Previous calculations, based on effective models of QCD, require 
  $|\thetabar| \leq 10^{-9}$ \cite{NEDM:th}.  
  The puzzle of explaining such an unnaturally small number is referred to as 
  the strong CP problem \cite{CP:review}.

  The problem is difficult and interesting, since an analytical calculation of
  low energy hadronic observables based on the QCD Lagrangian is not a trivial
  task.
  Furthermore, it turns out that an important property
  of QCD, namely -- chiral symmetry, is closely related to the strong CP 
  problem and imposes three stringent constraints on the possible breaking of
  P and T symmetries in QCD, which include: (1) the necessity of non--zero 
  quark masses \cite{CP:chiral sym1}, (2) spontaneous chiral symmetry 
  breaking \cite{CP:chiral sym2}, and (3) the $U_A(1)$ anomaly 
  \cite{CP:chiral sym3} \cite{CP:chan1}. 
  These symmetry constraints not only dictate the functional dependence of all
  CP violating observables on the QCD parameters, but also provide a dynamical
  suppression to the CP violating observables \cite{CP:chan1}.
 
  A natural solution of the strong CP problem can be obtained without invoking 
  a tiny $\thetabar$ parameter if the dynamical suppression is sufficient to 
  diminish the CP violating observables below the experimental upper bound.
  Thus, we need to face the challenge how to realize these constraints 
  explicitly in our calculations without employing a perturbative expansion on
  the $\thetabar$ parameter. Our calculations on the NEDM problem, which is the
  first one based on the quark--gluon degrees of freedom and the QCD Lagrangian
  with a $\thetabar$ parameter, will provide a critical answer to such an 
  interesting scenario for the strong CP problem.

  This paper is organized as follows:
  An introduction to the NEDM problem and the strong CP violation is given in
  section \ref{sec:y2}, where we also set up the notations used in this work.
  The hadronic and quark--gluon representations of the nucleon correlation
  functions ( NCF ) are discussed in section \ref{sec:y3} and section 
  \ref{sec:y4}, respectively. 
  The results obtained from both representations of the NCF are used to derive
  QCD sum rules, which are analyzed in section \ref{sec:y5} and section
  \ref{sec:y6}. 
  We conclude with a brief summary in section \ref{sec:y7}. 

    \section{Introduction}
    \label{sec:y2}


Before we discuss the QCD sum rule calculations of the nucleon electric dipole
moments \cite{QSR:nmm}, it is useful to clarify some issues related to the 
strong CP violation in QCD. 

\begin{enumerate}

 \item {\bf CP Violations in QCD}

    Up to dimension four, the most general QCD Lagrangian in four dimensional
    space-time, consistent with Lorentz invariance, hermiticity, gauge
    invariance, is
   \begin{eqnarray}
   \label{eq:chiral phases}
   {\mathcal{L}}_{QCD} &\equiv&   {\bar \psi} \m i {\!\not\! D} \psi
                  +  \massq  {\bar \psi} \m e^{ i \thetaq \gamma_5 } \psi
                  +  \frac{1}{4}  G^2 
                  +  \frac{ g_s^2 \thetag }{ 32 \pi^2 } G \tilde G   \\ 
     \mbox{where} \hspace{2.5cm}
   {\!\not\! D}  &\equiv& ( \partial_\mu + i g_s B_\mu^a \frac{ \lambda^a }{2} )
                                      \cdot \gamma^\mu                 \\
     \mbox{and}   \hspace{2cm}
     {\tilde G}_{\mu \nu} &\equiv& \frac{1}{2} \epsilon_{\mu \nu \alpha \beta}
                                   G^{\alpha \beta}, \hspace{1.5cm}
                                   \epsilon_{0123}=1
  \end{eqnarray}
   The meanings of various symbols are:
  \begin{eqnarray*}
      \psi                &:& \mbox{quark field }                           \\
     {\bar \psi}          &:& \mbox{Dirac adjoint of the quark field, }
                             {\bar \psi} \equiv \psi^\dagger \gamma_0       \\
      B_\mu^a             &:& \mbox{gluon field, } a = 1,..,8               \\
      \frac{\lambda^a}{2} &:& \mbox{generators of the color } SU(3)
                              \mbox{ gauge group, } a = 1,..,8               \\
      G_{\mu \nu}         &:& \mbox{gluonic tensor field, } 
                               G_{\mu \nu} \equiv
                             [\mbox{ } \partial_\mu + i g_s B_\mu, 
                              \mbox{ } \partial_\nu + i g_s B_\nu \mbox{ }],
                              \mbox{ } G^2 \equiv  G_{\mu \nu} G^{\mu \nu}  \\ 
      g_s                 &:& \mbox{strong coupling constant in QCD}        \\
      \thetaq             &:& \mbox{quark chiral phase}                     \\
      \thetag             &:& \mbox{gluon chiral phase}
  \end{eqnarray*}
    
   Here we have two P and CP violating ( but C even ) terms in the QCD 
   Lagrangian,
   $$i \massq \sin \thetaq \m {\bar \psi} \gamma_5 \psi, \hspace{1cm}
      \frac{ g_s^2 \thetag }{ 32 \pi^2 } G \tilde G $$
   the former is referred to as a quark pseudo--mass term, the second is 
   referred to as a gluon anomaly term. 
   Our notation is chosen such that the corresponding CP violating parameters
   $\thetaq$ and $\thetag$ are angular variables\footnote{The gluon anomaly
   term, when evaluated with instanton configurations, gives integer values.
   Thus, the QCD generating functional is a periodic function of $\thetag$.}
   and the QCD generating functional is periodic with respect to these CP 
   violating parameters.
   The general QCD Lagrangians with two chiral phases $\thetaq$, $\thetag$ are
   not all physically independent. 
   Through the $U_A(1)$ anomaly \cite{QCD:ABJ Anomaly} \cite{QCD:Fuji}, we can 
   shift some part of $\thetaq$ to $\thetag$, and vice versa, by performing an
   $U_A(1)$ rotation on the quark field  
   \begin{eqnarray}
    \psi &\rightarrow& \psi'         \equiv       e^{ i \theta \gamma_5 } \psi, 
     \hspace{1.87cm}  {\psi'}^\dagger = \psi^\dagger e^{ i \theta \gamma_5 }, \\
    \thetaq &\rightarrow& \thetaq - 2 \theta, \hspace{2.73cm}
    \thetag  \rightarrow  \thetag - 2 \theta.
   \end{eqnarray}
   Therefore, only the difference between these two phases,
   \begin{equation}
   \thetabar \equiv \thetag - \thetaq
   \end{equation}
   which is invariant under an $U_A(1)$ rotation, is a physical parameter and 
   can be used to label the equivalent classes of CP violating QCD Lagrangians
   \cite{CP:chan1}.
   Furthermore, since physical observables should be independent of the 
   reparameterization of the Lagrangian, we conclude that a CP violating 
   observable should only be proportional to $\thetabar$, instead of being 
   an arbitrary function of $\thetaq$ and/or $\thetag$.
   
   In addition to the explicit symmetry-breaking parameter $\thetabar$, there
   are three important symmetry constraints which could suppress the magnitude
   of a CP violating observable. 
   These constraints include: (1) explicit chiral symmetry breaking due to
   finite current quark masses, (2) spontaneous chiral symmetry breaking with a
   nonzero quark condensate, and (3) the $U_A(1)$ anomaly \cite{CP:chan1}.
   Suppression of the CP violating observables through the symmetry constraints
   is possible because in any of these particular limits, the effects of strong
   CP violation vanish, even with a nonzero $\thetabar$ parameter. 
   Therefore, it is desirable to have a calculation of a CP violating
   observable, e.g., $d_N$, without assuming a perturbative expansion in the 
   $\thetabar$ parameter. 
   Such a calculation could provide an answer to the important question: Why is
   strong CP violation small?

 \item {\bf Nucleon Electric Dipole Moment as an EM form factor}

   In a nondegenerate system like the neutron, the existence of an EDM implies
   the violation of both parity ( P ) and time reversal ( T, or CP ) symmetries.
   To establish a connection between the CP violating parameter $\thetabar$ in
   QCD and the NEDM, it is useful to study the nucleon EM matrix element:
  \begin{eqnarray}
   V_\mu^N ( q; p_1, p_2 ) & \equiv & 
      \ftr{q}{x} \m \langle N(p_2)|J_\mu (x)|N(p_1) \rangle  \nonumber \\
                 & =      & (2 \pi)^{2\omega} {\delta}^{2\omega} (q-p) \m
                    \langle N(p_2)|J_\mu (0)|N(p_1) \rangle  \nonumber \\
                 & \equiv & (2 \pi)^{2\omega} {\delta}^{2\omega} (q-p) \m
                            V_\mu^N (p_2,p_1)                          \\
     \mbox{ with } \hspace{1cm} p  & \equiv & p_2 - p_1 = q 
  \end{eqnarray}
   and $2\omega$ is the space--time dimension.
   The nucleon EM vertex is extracted from the nucleon EM matrix element by
   factoring out the ( on--shell ) Dirac spinor for the nucleon state:
  \begin{equation}
     V_\mu^N (p_2,p_1) = {\bar u}(p_2) \m \Gamma_\mu^N (p_2,p_1) \m u(p_1)
  \end{equation}
   Using (1) current conservation $\partial_\mu J^\mu = 0$ and (2) hermiticity
   $V_\mu^N (p_2,p_1) = [ \m V_\mu^N (p_2,p_1) \m ]^\dagger$, we can write 
   down a general form for the EM vertex $ \Gamma_\mu^N (p_2,p_1) $ of
   spin $1/2$ on-shell nucleon state:
  \begin{equation}
  \label{eq:em form factors}
   \Gamma_\mu^N (p_2,p_1)
  =     F_1^N (q^2) \gamma_\mu
    + i F_2^N (q^2) \frac{q^\nu \sigma_{\mu \nu}         }{2M_N}
    -   F_3^N (q^2) \frac{q^\nu \sigma_{\mu \nu} \gamma_5}{2M_N}
    +   F_4^N (q^2) (q^2 \gamma_\mu - q_\mu \hat{q} ) \gamma_5
  \end{equation}
   where we have four form factors $F_1^N, F_2^N, F_3^N,$ and $F_4^N$
   characterizing the EM properties of the nucleons.
   At $q^2=0$, they are the various EM moments\footnote{Due to the use of a 
   constant EM background, the contribution of the anapole moment $F_4^N$
   vanishes in our calculations.} of the nucleon state:
  \begin{eqnarray}
                   e   \m   F_1^N (q^2=0)
             &=&  Q_N  \m (\mbox{ charge                 })             \\
      \frac{e}{2M_N}   \m [ F_1^N (q^2=0) + F_2^N (q^2=0) ]
             &=& \mu_N \m (\mbox{ magnetic moment        })             \\
      \frac{e}{2M_N}   \m [ F_3^N (q^2=0) ]
             &=&  d_N  \m (\mbox{ electric dipole moment })             \\
      \frac{1}{2M_N^2} \m [ F_4^N (q^2=0) ]
             &=&  a_N  \m (\mbox{ anapole moment }) 
  \end{eqnarray}
   It is useful to notice that the tensor structure associated with the
   anomalous magnetic moment $F_2^N$ ( $\sigma_{\mu \nu}$ ) and that associated 
   with the electric dipole moment $F_3^N$ only differs by a factor
   $i \gamma_5$.
   In view of this, we can rewrite these EM form factors in a polar form,
   \begin{eqnarray}
   \label{eq:em moments}
    F_2^N + i F_3^N \gamma_5 &\equiv& F_N e^{i \alpha_N \gamma_5}      \\ 
                 {( F_N )}^2 &\equiv& {( F_2^N )}^2 + {( F_3^N )}^2    \\
               \tan \alpha_N &\equiv& \frac{F_3^N}{F_2^N}          
   \end{eqnarray}

 \item {\bf Calculation of Hadronic Matrix Elements from QCD}
          
   At first sight, to calculate nucleon EM moments from QCD requires a 
   knowledge of the nucleon wave function in terms of the quark gluon basis and
   a technique of solving the nontrivial strong coupling dynamics in the low
   energy region.
   This is certainly beyond our current ability ( except for numerical lattice
   calculations ) and we have to rely on other approaches to avoid these 
   complications. 
   For this purpose, we choose to calculate a nucleon correlation function
   ( NCF ) in the presence of external EM fields.
   \begin{equation}
    \Pi_N (p) \equiv \ftr{p}{x} 
    \varcond{ \m T \m \eta_N (x) \m {\bar \eta}_N (0) \m }
            { \thetaq, \thetag }{ F_{\mu \nu} }       
   \end{equation}
   where the nucleon interpolating field $\eta_N$ is a composite quark operator
   carrying the same quantum number as a nucleon. 
   For our calculation, we choose
   \begin{equation}
   \label{eq:nif}
      \eta_n \equiv ( d^t C \gamma_\mu d ) \gamma_5 \gamma^\mu u
   \end{equation}
   as a neutron interpolating field \cite{QSR:current}. 
   A similar expression for the proton can be obtained by exchanging u and d
   quarks.
   
   In this approach, the EM matrix element of nucleons, which describes the 
   response of the nucleon states to the weak external perturbation, can be
   imbedded in the first order expansion of the NCF with respect to the 
   external field.
   Since we shall focus on the EM form factors at zero momentum transfer,
   the EM fields can be taken as constants \cite{QSR:external field}.

   To the first order of the electric charge, the NCF $\Pi_N (p)$ can be
   expanded as
   \begin{equation}
    \Pi_N (p) \m \equiv \m \Pi^{(0)}_N (p) \m + \m e \Pi^{\mu \nu}_N (p)
                        F_{\mu \nu} \m + \m O( e^2 )
   \end{equation}
   We shall call $\Pi^{(0)}_N (p)$ a nucleon propagator and 
   $\Pi^{\mu \nu}_N (p)$ a ( nucleon ) polarization tensor. The former
   describes the propagation of hadronic states carrying the nucleon quantum
   numbers, the latter gives the EM vertices of hadronic states with the 
   external fields. 
   
 \item {\bf Tensor Structures of the Nucleon Correlation Function}

   It is important to know how to write down a complete set of covariant 
   tensors ( these will be referred to as basis tensors, composed of one 
   Lorentz vector $p_\mu$ and 16 Dirac matrices ) and decompose the NCF in
   terms of these basis tensors. 
   Such structures come out naturally from both hadronic representations 
   ( see the discussions in Sec.~\ref{sec:y3} ) and QCD calculations ( see the
   discussions in Sec.~\ref{sec:y4} ) for the NCF. 
   Furthermore, the QCD sum rules are extracted from the coefficient functions
   associated with these basis tensors.
   A correct decomposition will help assure that there are no omissions and
   redundancies in our calculations.

   We find that it is convenient to use a commutation-anticommutation relation
   analysis \cite{math:tensor trick} to generate all possible independent 
   invariant tensors for the NCF.
   For the nucleon propagator $\Pi_N^{(0)}(p)$, there are 4 independent 
   tensors:
   \begin{equation}
    I,\hspace{0.5cm} \gamma_5,\hspace{0.5cm}
    \hat{p} \equiv p_\mu \cdot \gamma^\mu,
    \hspace{0.5cm} \hat{p} \gamma_5.
   \end{equation}
   For the polarization tensor $\Pi_{N}^{\mu \nu}(p)$, the tensor basis
   consists of 8 independent second rank tensor matrices:
   \begin{eqnarray}
             \sigma^{\mu \nu}&,&
             \hspace{1cm}
             \sigma^{\mu \nu} \cdot \gamma_5                               \\
             p^\mu \gamma^\nu - p^\nu \gamma^\mu&,&
             \hspace{1cm}
           ( p^\mu \gamma^\nu - p^\nu \gamma^\mu ) \cdot \gamma_5          \\
     \epsilon^{\mu \nu \alpha \beta} p_\alpha \gamma_\beta&,&
     \hspace{1cm}    
     \epsilon^{\mu \nu \alpha \beta} p_\alpha \gamma_\beta \cdot \gamma_5  \\
   \hat{p} ( p^\mu \gamma^\nu - p^\nu \gamma^\mu )&,&
   \hspace{1cm}
   \hat{p} ( p^\mu \gamma^\nu - p^\nu \gamma^\mu ) \cdot \gamma_5
   \end{eqnarray}
   Given these classifications of the basis tensors for the NCFs, we can 
   define various coefficient functions associated with them.
\begin{eqnarray}
      \Pi_N (p) & \m \equiv \m & \Pi_N^{(0)}(p) 
                  \m +      \m e \Pi_N^{\mu \nu}(p) F_{\mu \nu}
                  \m +      \m O(e^2)                                       \\
                &              &                                  \nonumber \\
 \label{eq:polar1a}
 \Pi_N^{(0)}(p) & \m \equiv \m &          f_1^N (p^2) \cdot \hat{p}
                  \m +      \m   {\tilde f}_2^N (p^2) \cdot I
                  \m +      \m i {\tilde f}_3^N (p^2) \cdot \gamma_5        \\
 \label{eq:polar1b}             
                & \m \equiv \m &          f_1^N (p^2) \cdot \hat{p}
                  \m +      \m            f_2^N (p^2) \cdot 
                                    e^{i \phi_N (p^2) \gamma_5}             
\end{eqnarray}
\begin{eqnarray} 
   \mbox{with} \hspace{1.5cm}
   [f_2^N (p^2)]^2   & \equiv &      {[{\tilde f}_2^N (p^2)]}^2
                              +      {[{\tilde f}_3^N (p^2)]}^2             \\
   \tan \phi_N (p^2) & \equiv & \frac{ {\tilde f}_3^N (p^2) } 
                                     { {\tilde f}_2^N (p^2) }
\end{eqnarray}
 \begin{equation}
 \label{eq:polar2}
  \begin{array}{cclcl}
   \Pi_N^{\mu \nu}(p) &\equiv&
        {\tilde g}_1^N (p^2) \m   \sigma^{\mu \nu}
  &+&   {\tilde g}_2^N (p^2) \m i \sigma^{\mu \nu} \gamma_5               \\
  &+&            g_3^N (p^2) \m i \epsilon^{\mu \nu \alpha \beta}
                                   p_\alpha \gamma_\beta \gamma_5
  &+&    g_4^N (p^2) \m         i (p^\mu \gamma^\nu - p^\nu \gamma^\mu) 
                                         \gamma_5                         \\
  &+&    g_5^N (p^2) \m   \hat{p} (p^\mu \gamma^\nu - p^\nu \gamma^\mu)
  &+&    g_6^N (p^2) \m i \hat{p} (p^\mu \gamma^\nu - p^\nu \gamma^\mu)
                                         \gamma_5                         \\
  & &                                                                     \\
  &\equiv&
      g_1^N (p^2) \m \sigma^{\mu \nu} e^{i \varphi_1^N (p^2) \gamma_5} && \\
  &+& g_3^N (p^2) \m i \epsilon^{\mu \nu \alpha \beta}
                        p_\alpha \gamma_\beta \gamma_5
  &+& g_4^N (p^2) \m i (p^\mu \gamma^\nu - p^\nu \gamma^\mu) \gamma_5   \\
  &+& g_2^N (p^2) \m \hat{p} ( p^\mu \gamma^\nu - p^\nu \gamma^\mu )
                             e^{i \varphi_2^N (p^2) \gamma_5} && 
  \end{array} 
 \end{equation}
 \begin{eqnarray}
  \mbox{with} \hspace{3cm}
  [g_1^N (p^2)]^2        &\equiv&       [{\tilde g}_1^N ( p^2 )]^2  
                                +       [{\tilde g}_2^N ( p^2 )]^2    \\
  \tan \varphi_1^N (p^2) &\equiv&
          \frac {{\tilde g}_2^N ( p^2 )}{{\tilde g}_1^N ( p^2 )}      \\
  {[g_2^N (p^2)]}^2      &\equiv&     {[g_5^N (p^2)]}^2  + {[g_6^N (p^2)]}^2 \\
  \tan \varphi_2^N (p^2) &\equiv& \frac { g_6^N (p^2) }{ g_5^N (p^2) }
 \end{eqnarray}
  Notice that because of charge conjugation symmetry, the coefficient function
  associated with the $\hat{p} \gamma_5$ tensor in the nucleon propagator and
  those associated with the $p^\mu \gamma^\nu - p^\nu \gamma^\mu$ and
  $\epsilon^{\mu \nu \alpha \beta} p_\alpha \gamma_\beta$ tensors are
  identically zero.

  The discussion in the next two sections will be devoted to the constructions
  of a hadronic and a quark-gluon parameterizations for all these invariant 
  coefficients.

 \item {\bf CP Violating Vacuum Condensates and the Quark-Gluon Chiral Phases}

   In the QCD sum rule \cite{QSR:svz} calculations of the hadron observables,
   we use an operator product expansion ( OPE ) to expand a hadronic
   correlation function in the order of operator dimensions. The
   nonperturbative ( to be precise, nonanalytic in the strong coupling
   constant ) structure of the QCD dynamics is parameterized in terms of
   various vacuum condensates and the perturbative contributions ( Wilson
   coefficients ) can be calculated using Feynman rules.
   By truncating the OPE series at a certain dimension, we obtain an 
   approximate representation of the hadronic correlation function in terms of
   QCD parameters.
   On the other hand, the hadronic observables can be built in the hadronic
   correlation function by inserting a complete hadronic states and expanding
   the hadronic correlation function according to the hadron invariant masses.
   Through a matching between quark-gluon and hadron
   representations, the values of ground state observables can be extracted.
   It is crucial that we obtain the values of various vacuum condensates from
   other sources rather than calculating them within the sum rule method. 
   This becomes a problem if we need to include higher dimensional 
   condensates, which are poorly known. 
   Also, in the external field method, due to the polarization of the QCD 
   vacuum, there appear so-called induced condensates we need to take into 
   account \cite{QSR:nmm}.
   If CP is not a good symmetry of QCD, there could be in principle more unknown
   condensates associated with the CP violating operators, e.g., 
   $i \cond{ \bar q \gamma_5 q }$ and $\cond{ G \tilde G }$ . 
   However, it is possible to relate these parity doublet condensates
   ( $\cond{ \bar q q }$ and $i \cond{ \bar q \gamma_5 q }$ )
   to the chiral phases we have introduced.
   Indeed, there is a simple theorem, which can be used to relate quark 
   condensates in the parity doublets:
    \begin{theorem}
     \label{th:cct}
     $$ R_q^2 \equiv \left[   \cond{ \bar q          q } \right]^2 +
                     \left[ i \cond{ \bar q \gamma_5 q } \right]^2
     \m \m \mbox{is invariant under $U_A(1)$ chiral rotations.} $$
    \end{theorem}
  
   The above theorem implies that the two real number $\cond{ \bar q q }$ and
   $i \cond{ \bar q \gamma_5 q }$ can be thought of as the coordinates of a two
   dimensional plane, with $R_q$ defining the radius of a chiral circle
   generated by the $U_A(1)$ chiral rotations.
   \begin{eqnarray}
   \label{eq:chiral rad1}
       \cond{ \bar q          q } &\equiv& - R_q \cos \thetag   \\
   \label{eq:chiral rad2}
     i \cond{ \bar q \gamma_5 q } &\equiv& - R_q \sin \thetag  
   \end{eqnarray} 
   For this reason, we shall refer $R_q$ as a chiral radius\footnote{The value
   of a chiral radius depends on the quark mass.
   For light flavor $m_q \leq \Lambda_{QCD}$, spontaneous chiral symmetry
   breaking implies $R_q$ is a finite positive number in the massless limit.}
   \footnote{For more than one light quark flavor, the values of the chiral
    phase 
   $$ \thetagq \equiv \arctan \frac{ i \cond{ \bar q \gamma_5 q } }
                                   {   \cond{ \bar q          q } } $$ can be
    determined from Crewther's condition \cite{QSR:chiral sym1}
   \begin{eqnarray}
   \label{eq:crewther1}
    \sum_q \thetagq &=& \thetag   \\
   \label{eq:crewther2}
    m_{q_i} R_{q_i} \sin ({\thetaq_i} - {\thetagq_i}) &=&
    m_{q_j} R_{q_j} \sin ({\thetaq_j} - {\thetagq_j})
   \end{eqnarray} }.
   Further use of this theorem will be discussed in latter section,
   see Sec.\ref{sec:y4}.
 
 \item {\bf $U_A(1)$ Chiral Rotations and the Use of the Polar Form in the Sum
            Rule Calculation}

   We have seen in the previous discussions that we can rewrite many variables
   and/or parameters in the polar form. These include:
   \begin{enumerate}
    \item the quark ( $\thetaq$ ) and gluon ( $\thetag$ ) chiral phases in the 
          general CP violating QCD Lagrangian ( Eq.(\ref{eq:chiral phases}) ),
    \item the anomalous magnetic moment ( $F_2^N$ ) and the electric dipole
          moment ( $F_3^N$ ) of the nucleon ( Eq.(\ref{eq:em moments}) ),
    \item the invariant coefficient functions associated with the tensor basis
          ( Eqs.(\ref{eq:polar1a}),(\ref{eq:polar1b}),(\ref{eq:polar2}) ),
    \item the quark condensates in the parity doublet 
          ( Eqs.(\ref{eq:chiral rad1}),(\ref{eq:chiral rad2}) ).
   \end{enumerate}

   We find it is quite convenient to adopt this convention for the following
   reasons:
   \begin{enumerate}
    \item The reparameterization invariance of physical observables under the
          $U_A(1)$ chiral transformations of the QCD Lagrangian can be 
          maintained throughout our calculation. 
          Thus, there is no need to stick to a particular representation of
          the QCD Lagrangian.

    \item The polar form allows natural identifications among chirally invariant
          and/or chirally covariant variables. 
          Those chirally invariant observables depend only on the chirally 
          invariant parameters, e.g., $m_q$, $R_q$, and $\thetabar$; the 
          chirally covariant variables change by a constant phase under 
          $U_A(1)$ chiral rotations.

    \item Since the $\thetabar$ parameter appears as an angular variable, one
          can solve the $\thetabar$ dependence of the CP violating observables
          without using a perturbative expansion. Furthermore, the periodic
          structure of the $\thetabar$ dependence comes out automatically due
          to the polar form.

    \item The symmetry constraints on the strong CP violations in QCD can be 
          made transparent in the sum rule relations if we organize both the
          OPE series and the hadronic representations in the polar forms.
          Without an explicit solution to the sum rule relations, one can show
          that CP violating observables vanish if chiral symmetry is exact in
          QCD.
   \end{enumerate}

\end{enumerate}

    \section{Study of the Nucleon Correlation Function ( NCF ) 
             from the Hadron Degrees of Freedom}
    \label{sec:y3}


\begin{enumerate}

\item {\bf Basics}

     One advantage of the QCD sum rule calculation is to extract ground state 
     observables from a correlation function without knowing the exact wave 
     function of the nucleon state.
     The price for this convenience is that the interpolating field we choose
     couples to all possible hadronic excited states with the same quantum 
     number as the nucleon. 
     In addition, the ground state matrix element we are interested in is often
     accompanied with other excited state contributions to the correlation 
     function.
     Consequently, the extraction of ground state observables from a NCF is
     possible only if we can identify and isolate various contributions to the
     NCF from hadronic states.
     To achieve this purpose, we insert a complete set of hadronic states in
     the NCF $\Pi_N (p)$ of the interpolating field $\eta_N$. 
     In doing so, we can factorize the correlation function into nucleon
     spinors and hadronic matrix elements.
     While the ground state observables, e.g., mass $M_N$ and EM moments
     $F^N_i$ of a nucleon $N$, can be specified explicitly, the fine details of
     the excited state spectrum are smeared out by employing suitable
     parameterizations \cite{QSR:external field}.
     The ground state nucleon observables, together with the excited state
     parameters are basic ingredients of a hadronic representation of the NCF. 
     We shall discuss the constructions of hadronic representations of the 
     nucleon propagator and the polarization tensor in the following two 
     subsections.

     Since we wish to maintain an $U_A(1)$ reparametrization covariance in our 
     calculations, it is important to keep all chiral phases explicit.
     In particular, we fix the chiral phases of the physical hadron states 
     ( and the QCD vacuum, denoted as $\Omega$ ) to be zero and allow quark 
     fields to be in any chiral basis.
     With this point in mind, the definition of the nucleon spinor is given by
   \begin{equation}
   \label{eq:nucleon spinor}
    \cond{ \m \Omega \m | \m \eta_N \m | \m N( \vec{p}, s_N ) \m } \equiv 
    \lambda_N \m e^{ i \frac{ \theta_N }{ 2 } \gamma_5 } \m u( \vec{p}, s_N )
   \end{equation}
     Here the nucleon residue $\lambda_N$ gives the overlap between the nucleon
     state $N$ and the interpolating field $\eta_N$, the nucleon chiral phase
     $\theta_N$ specifies the quark basis of the QCD Lagrangian.
     It is convenient to choose an interpolating field which transforms like a
     quark field under an $U_A(1)$ rotation. 
     For instance, for our chosen neutron interpolating field $\eta_n$
     ( Eq.(\ref{eq:nif}) )
     \begin{eqnarray}
           u  &\rightarrow&        u' = e^{i \frac{\thetau}{2} \gamma_5} u,
               \hspace{2cm}
           d   \rightarrow         d' = e^{i \frac{\thetad}{2} \gamma_5} d    \\
      \eta_n  &\rightarrow&  \eta'_n  = e^{i \frac{\thetau}{2} \gamma_5} \eta_n,
               \hspace{1.6cm} 
      \theta_n \rightarrow \theta'_n  = \theta_n + \thetau
     \end{eqnarray}
     In this case, the nucleon chiral phase $\theta_N$ transforms covariantly 
     ( changing by a constant phase ) 
     and the nucleon residue $\lambda_N$ stays invariant under an $U_A (1)$ 
     rotation.

\item {\bf Hadronic Representation of the Nucleon Propagator}

     For the nucleon propagator $\Pi^{(0)}_N (p)$, we can insert a complete set
     of hadronic states $\sum_N |N \rangle \langle N| = 1$ between the
     time-ordered product of the interpolating fields $\eta_N, {\bar\eta}_N$. 
 \begin{eqnarray}
     \Pi_N^{(0)}(p) &\equiv&          \ftr{p}{x} \m
     \cond{ \Omega  | T ( \eta_N (x), \bar{\eta}_N (0) ) | \Omega  }  \\
                      &\equiv& \sum_{N} \ftr{p}{x} \m \theta( x_0)
     \cond{ \Omega  | \eta_{N}(x)                        | N       } 
     \cond{ N       | \bar{\eta}_N (0)                   | \Omega  }
     \nonumber \\            & & \hspace{1.9cm}    - \m \theta(-x_0)
     \cond{ \Omega  | \bar{\eta}_N (0)                   | \bar{N} }
     \cond{ \bar{N} | \eta_{N}(x)                        | \Omega  }
 \end{eqnarray}
     With the definition of the nucleon spinor in a general quark chiral basis 
     ( Eq.(\ref{eq:nucleon spinor}) ) and the standard procedure to simplify
     the algebra, we obtain a propagator of the nucleon state, with an overall
     chiral conjugation\footnote{A chiral conjugation of a two--point Green's
     function is defined as $ G ( p ) \rightarrow U ( \theta ) \cdot G ( p )
     \cdot U ( \theta ) $, with $ U ( \theta ) = e^{ i \frac { \theta } {2} 
     \gamma_5 } $.}:
 \begin{equation}
 \label{eq:nuprop1}
     e^{ i \frac {  \theta_N } {2} \gamma_5 }       \m
           \frac { \lambda_N^2 } { \hat{p} - M_N }  \m
     e^{ i \frac {  \theta_N } {2} \gamma_5 }       \m = \m
                   \lambda_N^2                      \m
     \left( \m \frac { \hat{p} + M_N \cdot e^{ i \theta_N \gamma_5 } } 
                     { p^2 - M_N^2 }                \m \right)
 \end{equation}
     Similar expressions can be written for excited states $N^*$ with 
     different residues $\lambda_{N^*}$ and total masses $M_{N^*}$.
       
     The contributions of the excited states to the NCF lead to many unknowns
     in our calculations\footnote{This problem can be traced back to the choice
     of an interpolating field for the nucleon state. A perfect choice of an
     interpolating field is equivalent to the solution of an exact wave
     function of the nucleon state, which couples only to the nucleon state and
     has zero overlap with any excited state.}. 
     Consequently, we shall take a simple parameterization to replace the
     contributions of all excited states without involving the complete 
     hadronic spectrum.
     Based on a duality argument, we can identify the total contributions of 
     the excited states to the nucleon propagator as the the leading terms from
     the quark-gluon calculations, starting from a continuum threshold $s_0^N$:
     \begin{equation}
     \label{eq:nuprop2}
              Re \m f_i^N (p^2) \m ( \mbox{ continuum } )
      =\frac{1}{\pi} \m Pr. \int_{s_0^N}^\infty  
       \frac{ Im \m f_i^N (s) \m ( \mbox{ quark-gluon } ) }{ p^2 - s } \m ds
     \end{equation}

     Here we use a dispersion relation for the invariant coefficient functions
     $f_i^N$ ( See Eqs.(\ref{eq:polar1a}),(\ref{eq:polar1b}),(\ref{eq:polar2})
     ) to relate the contributions from two representations and $Pr.$ means
     principal value of the complex integral.
     The calculations of the quark-gluon representation of the NCF will be
     given in the next section.

     The hadronic representation of the nucleon propagator $\Pi^{(0)}_N (p)$ is
     given by the sum of Eq.(\ref{eq:nuprop1}) and Eq.(\ref{eq:nuprop2}),
     where we have three nucleon variables, $\lambda_N, \theta_N$, and $M_N$
     to be determined from the QCD sum rules.
                                          
\item {\bf Hadronic Representation of the Polarization Tensor}

     Since the polarization tensor $\Pi^{\mu \nu}_N (p)$ comes from a
     time-ordered product of the electromagnetic current $J_\mu$ and the 
     interpolating fields $\eta_N, {\bar\eta}_N$, we insert two complete sets
     of hadron states, 
   $\sum_N |N \rangle \langle N| = \sum_{N'} |{N'} \rangle \langle {N'}| = 1$. 
\begin{eqnarray}
      e \Pi^{\mu \nu}_N (p) F_{\mu \nu} &\equiv& 
     \ftr{p}{x} \int d^{2 \omega} z \m
     \cond{ \m T {\mathcal{L}}_{int} (z) \eta_N (x) \bar{\eta}_N (0) \m } \\
     &=& \sum_N \m \sum_{N'} \m \ftr{p}{x} \int d^{2 \omega} z \m
         \theta(x-z) \m \theta(z-0) \times  \nonumber \\
          &\times& \cond{ \Omega | \eta_N (x)      | N      } 
                   \cond{ N      | J_{\mu} (z)     | N'     }
                   \cond{ N'     |\bar{\eta}_N (0) | \Omega } A^\mu (z) 
           +       \nonumber \\
          &+&      \mbox{time ordering}
\end  {eqnarray}
     where the vector potential $A^\mu$ of a constant EM field $F_{\mu \nu}$
     is given by
\begin{equation}
\label{eq:em tensor}
     A^\alpha \equiv \frac{1}{2} x_\beta F^{\beta \alpha}
\end{equation}
     The two insertions are independent. Hence, the polarization tensor 
     $\Pi^{\mu \nu}_N (p)$ contains, in addition to the (1) nucleon ground 
     state EM form factors, (2) ground state to excited states transitions, and
     (3) transitions among excited states\footnote{The transitions in a 
     constant EM background are possible because we are looking at highly
     off-shell states.}. 
     We can think of the polarization tensor, in its hadronic representation, 
     as a huge matrix in the Fock space and the three contributions are given
     in a tabular form ( see Table 1 ):

     We shall discuss these contributions to various invariant coefficient
     functions of the polarization tensor, with the superscript labels
     $g_{Ni}^{(1)}, g_{Ni}^{(2)}, g_{Ni}^{(3)}$ ( $i$ is the index for the
     independent tensor basis ) corresponding to three regions listed above.
     Thus,
     \begin{equation}
       g_i^N = g_{Ni}^{(1)} + g_{Ni}^{(2)} + g_{Ni}^{(3)},
       \hspace{1.5cm} i=1,2,...,6
     \end{equation}             

     \begin{enumerate}
     
      \item the nucleon ground state EM form factors

            If we only take the ground state nucleon from the double sum over
            hadronic complete sets, the polarization tensor reduces to a 
            product of two ( CP conserving ) nucleon propagators and the
            nucleon EM vertex, with an overall $U_A(1)$ chiral conjugation.
            \begin{eqnarray}
            \label{eq:polart1}
            & & e \m \Pi^{\mu \nu}_N (p) ( \mbox{ nucleon state } ) 
                  \m F_{\mu \nu} \nonumber \\
            &=& \int d^{2 \omega} z \m
            e^{ i \frac { \theta_N } {2} \gamma_5 }  \m \left[ \m
            \frac { \lambda_N   } { \hat{p} - M_N }  \m
            \cond{ N     | J_{\mu} (z)     | N    }  \m
            A^\mu (z)                                \m
            \frac { \lambda_N^* } { \hat{p} - M_N }  \m \right]
            e^{ i \frac { \theta_N } {2} \gamma_5 }  \m
            \nonumber \\
            &+& \mbox{time ordering}
            \end{eqnarray}
            Substituting all ingredients ( Eqs.(\ref{eq:em form factors}),
            (\ref{eq:nucleon spinor}), (\ref{eq:em tensor}) ) into the equation 
            Eq.(\ref{eq:polart1}), we obtain the contribution from the ground
            state nucleon to the polarization tensor ( We have factored out an
            overall constant
            $\frac{ \lambda_N^2 F_{\mu \nu} }{ 4M_N {(p^2- M_N^2)}^2 }$ ):
\begin{eqnarray}
  \label{eq:polara1}
            g_{N1}^{(1)} (p) &=& 2 i  \cos \theta_N        M^2_N  F_1^N
                                 + i  \cos \theta_N (p^2 + M^2_N) F_2^N
                                 -    \sin \theta_N (p^2 - M^2_N) F_3^N     \\
            g_{N2}^{(1)} (p) &=& 2 i  \sin \theta_N        M^2_N  F_1^N
                                 + i  \sin \theta_N (p^2 + M^2_N) F_2^N
                                 +    \cos \theta_N (p^2 - M^2_N) F_3^N     \\
            g_{N3}^{(1)} (p) &=& 2 M_N ( F_1^N + F_2^N )                    \\
            g_{N4}^{(1)} (p) &=& 2 M_N   F_3^N                              \\
            g_{N5}^{(1)} (p) &=&(-2) \m   [ \m \cos   \theta_N F_2^N
                                             + \sin   \theta_N F_3^N \m ]
                              = (-2) \m F_N \m \cos ( \theta_N - \alpha_N ) \\
  \label{eq:polara2}           
            g_{N6}^{(1)} (p) &=&(-2) \m   [ \m \sin   \theta_N F_2^N
                                             + \cos   \theta_N F_3^N \m ]    
                              = (-2) \m F_N \m \sin ( \theta_N - \alpha_N )    
\end{eqnarray}
      \item ground state transitions to excited states
  
            As in the case of excited state contributions to the nucleon 
            propagator, we need to sum over all ground state to excited states
            transitions, and parameterize the spectral function with a few
            constants. This can be achieved if we take the following
            expression:
      \begin{eqnarray}
     & &       \varcond{ \m \Omega \m | \m \eta_N (0) \m | \m N \m }
                       { \thetaq, \thetag }
                       { F_{\mu \nu} \not= 0 } -
               \varcond{ \m \Omega \m | \m \eta_N (0) \m | \m N \m }
                       { \thetaq, \thetag }
                       { F_{\mu \nu}     = 0 }             \nonumber  \\
     &\equiv&   \int d^{2 \omega} z \m
                  \cond{ \m \Omega \m 
                       | \m T \m ( {\mathcal{L}}_{int} (z) \m \eta_N (0) )\m |
                         \m N ( \vec {p}, s_N ) \m } 
                       - \mbox{ pole term }                           \\
     \label{eq:excited constant}
     &\equiv&    e F_{\mu \nu} \m
       \left[ \m       E_N^A       \m e^{ i \frac{\varphi_N^A}{2} \gamma_5 }
                   \m   \sigma^{\mu \nu}                      \m 
        + \m \frac{E_N^B}{M_N} \m e^{ i \frac{\varphi_N^B}{2} \gamma_5 }
                   \m ( \m p^\mu \gamma^\nu - p^\nu \gamma^\mu \m ) \m \right]
          \m u ( \m \vec {p}, s_N \m ) 
      \end{eqnarray}
         This is the most general form for a nucleon spinor in the presence of
         an external EM field; we use the Dirac equation for the ( on shell ) 
         nucleon spinor $u (\vec {p}, s_N)$ to reduce the invariant tensor 
         structure with four unknown model parameters:
         $E_N^A, \varphi_N^A, E_N^B, \varphi_N^B$. 
         All of these parameters are invariant functions of $Q^2 \equiv - p^2$.
         The independent tensors are chosen such that $E_N^A,E_N^B$ are 
         invariant under an $U_A(1)$ rotation, and the phases $\varphi_N^A,
         \varphi_N^B$ transform covariantly. 
         As a crude approximation, we shall neglect the $Q^2$ dependence of
         these model parameters and treat them as constants in our
         calculation\footnote{A more detailed parameterization including the 
         nucleon pion continuum, which is of importance in terms of both chiral
         symmetry \cite{QSR:chiral sym2} and quantitative error analysis
         \cite{QSR:excitation}, can be used to check the
         validity of this approximation \cite{QSR:chan3}.}.

         The contributions to the polarization tensor of transitions from the
         nucleon to excited states is:
\begin{eqnarray}                                                        
       &&\sum_{N'} e \m \Pi_N^{\mu \nu}(p) ( N \rightarrow N' ) \m F_{\mu \nu}
         \equiv \ftr{p}{x}  \nonumber \\
       & &   \theta(x_0) \m
   [ \m \varcond{\Omega|\eta_N (x)|N}     {\thetabar}{ F_{\mu \nu} \not= 0}
\m - \m \varcond{\Omega|\eta_N (x)|N}     {\thetabar}{ F_{\mu \nu}     = 0} \m ]
     \m \varcond{N|\bar{\eta}_N(0)|\Omega}{\thetabar}{ F_{\mu \nu}=0 } + 
        \nonumber \\
       & & + \theta(x_0)
     \m \varcond{\Omega|\eta_N (x)|N}     {\thetabar}{ F_{\mu \nu}=0 } \m
   [ \m \varcond{N|\bar{\eta}_N(0)|\Omega}{\thetabar}{ F_{\mu \nu} \not= 0}\m  
\m - \m \varcond{N|\bar{\eta}_N(0)|\Omega}{\thetabar}{ F_{\mu \nu}     = 0}\m ]
     \m -   \nonumber \\
   & &  - \theta(-x_0) \m
  [ \m \varcond{\Omega|\bar{\eta}_N(0)|\bar N}{\thetabar}{ F_{\mu \nu} \not= 0}
\m -\m \varcond{\Omega|\bar{\eta}_N(0)|\bar N}{\thetabar}{ F_{\mu \nu} = 0} ]
    \m \varcond{\bar N|\eta_N (x)|\Omega}     {\thetabar}{ F_{\mu \nu}=0 } -
       \nonumber \\
   & &   -  \theta(-x_0)
    \m \varcond{\Omega|\bar{\eta}_N(0)|\bar N}{\thetabar}{ F_{\mu \nu}=0 } \m
  [ \m \varcond{\bar N|\eta_N (x)|\Omega}     {\thetabar}{ F_{\mu \nu} \not= 0}
\m- \m \varcond{\bar N|\eta_N (x)|\Omega}     {\thetabar}{ F_{\mu \nu}     = 0}
    \m ]
     \end{eqnarray}

       After collecting terms for each independent tensor basis of the
       polarization tensor, we obtain ( factoring out an overall constant
       tensor $\frac{ e \lambda_N^2 F_{\mu \nu} }{ 4(p^2 - M_N^2) }$ )
  \begin{eqnarray}
   \label{eq:polarb1}
      g_{N1}^{(2)} (p) &=&  2 E_N^A M_N  \cos( \frac {\varphi_N^A} {2} 
                                             + \frac {\theta_N}    {2} )   \\
      g_{N2}^{(2)} (p) &=&  2 E_N^A M_N  \sin( \frac {\varphi_N^A} {2} 
                                             + \frac {\theta_N}    {2} )   \\
      g_{N3}^{(2)} (p) &=& -2i     E_N^A \cos( \frac {\varphi_N^A} {2} 
                                             - \frac {\theta_N}    {2} )   \\
      g_{N4}^{(2)} (p) &=& 2i \m [ \m E_N^A \sin( \frac {\varphi_N^A} {2} 
                                                - \frac {\theta_N}    {2} )
                                    - E_N^B \sin( \frac {\varphi_N^B} {2} 
                                                - \frac {\theta_N}{2} ) \m ] \\
      g_{N5}^{(2)} (p) &=& 2i \m \frac{E_N^B}{M_N}
                                           \cos( \frac {\varphi_N^B} {2} 
                                               + \frac {\theta_N}    {2} )   \\
   \label{eq:polarb2}  
        g_{N6}^{(2)} (p) &=& 2i \m \frac{E_N^B}{M_N}
                                           \sin( \frac {\varphi_N^B} {2} 
                                               + \frac {\theta_N}    {2} ) 
  \end{eqnarray}

      \item transitions among excited states

            We shall apply the duality ansatz to identify the contributions to
            the polarization tensor of transitions among excited states; the
            leading contributions from the quark-gluon representation, starting
            from the continuum threshold $s_0^N$, are given by
            \begin{equation}
            \label{eq:polarc}
                     Re \m g_{Ni}^{(3)} (p^2) \m ( \mbox{ continuum } )
            = \frac{1}{\pi} \m Pr. \int_{s_0^N}^\infty                           
              \frac{ Im \m g_{Ni}^{(3)} (s)   \m ( \mbox{ quark-gluon } ) }
                   { p^2 - s } \m ds
            \end{equation}
            The procedure is similar to the case of the nucleon propagator. The
            quark-gluon calculation will be discussed in the next section.
   \end{enumerate}

     With all these ingredients, the hadronic representation of the 
     polarization tensor $\Pi^{\mu \nu}_N (p)$ is given by the sum of Eqs.
     (\ref{eq:polara1}) $\sim$ (\ref{eq:polara2}), (\ref{eq:polarb1}) $\sim$
     (\ref{eq:polarb2}), and (\ref{eq:polarc}), where we have two nucleon
     variables, $F_2^N, F_3^N$ ( or equivalently, $F_N, \alpha_N$ ) and four
     excited state unknowns, $E_N^A, \varphi_N^A, E_N^B, \varphi_N^B$ which
     will appear in the phenomenological side of the QCD sum rules.

\end{enumerate}

    \section{Study of the Nucleon Correlation Function ( NCF )
             from the Quark-Gluon Degrees of Freedom}
    \label{sec:y4}
	

\begin{enumerate}

 \item {\bf Basics}

       To represent the NCF in terms of the quark--gluon parameters,
       we use the method of operator product expansion ( OPE ) to calculate the
       NCF in QCD \cite{QSR:ope}.
       Because the NCF is the vacuum expectation value of composite quark
       operators, the expansion series of the NCF is a short distance expansion
       in the coordinate space, or, a ${\Lambda_{QCD}^2}/{Q^2}$ expansion in 
       momentum space\footnote{This is in contrast to the use of OPE in the
       deep inelastic scattering ( DIS ) of lepton off nucleon target. Where 
       because of the kinematic ( Bjorken limit ) and the state of interest 
       ( nucleon with a four--momentum $p_\mu$ ), the expansion series of the
       correlator in the DIS calculation is a light--cone expansion.}.
       We truncate the OPE series at dimension six, and the Wilson coefficients
       are calculated, using diagrammatic rules, in the first order of quark 
       mass $m_q$ and strong coupling constant $g_s$.

       A new feature associated with our problem is the presence of CP
       violating operators in the OPE series of the NCF. 
       For each CP conserving operator, there exists a parity partner which 
       coming from a $\gamma_5$ insertion or a dual transformation, e.g.,
       $\cond{\bar q q}$ vs. $\cond{\bar q \gamma_5 q}$ and $\cond{G^2}$ vs. 
       $\cond{G \tilde G}$.
       This is also true in the case of induced condensates, e.g.,
       $\cond{\bar q G_{\mu \nu} q}$ vs. $\cond{\bar q G_{\mu \nu} \gamma_5 q}$.
       These CP violating condensates introduce new unknowns in the microscopic
       representation of the NCF.
       Therefore, it is important that we know how to evaluate these new 
       condensates such that our calculation has predictive power.
       
       We can classify the CP violating condensates into two classes: the first
       class consists of $U_A(1)$ chiral covariant operators, e.g.,
       $i \cond{ \bar q \gamma_5 q }$, whose values depend
       on the representation of the QCD Lagrangian; the second class consists
       of $U_A(1)$ chiral invariant operators, e.g., $\cond{G \tilde G}$
       and $\cond{ \bar q \gamma_\mu q  \bar q \gamma^\mu \gamma_5 q}$. 
       The classification is useful because there is a general relation which 
       connects the parity partners in the first class condensates. 
       In particular, one can prove the generalized chiral circle theorem:
\begin{theorem}
$$  {\left[ \m   \cond{ \bar q \Gamma f ( G_{\mu\nu} )          q } \m \right]}^2 
  + {\left[ \m i \cond{ \bar q \Gamma f ( G_{\mu\nu} ) \gamma_5 q } \m \right]}^2 
       \mbox{ is invariant under $U_A(1)$ chiral rotations } $$
\end{theorem}
 
       Compared with the simplest case in Theorem 1, here $\Gamma$
       is an arbitrary Dirac matrix and $f ( G_{\mu\nu} )$ stands for any
       gauge invariant function of the gluon field tensor.

       Use of the polar representation for quark condensates, see 
       Eqs.(\ref{eq:chiral rad1}),(\ref{eq:chiral rad2}) allows the combination
       of $\cond{ \bar q q} + \cond{ \bar q \gamma_5 q} \gamma_5$ to be written
       as 
       \begin{equation} 
       \cond{ \bar q q } + \cond{ \bar q \gamma_5 q } \gamma_5 
       = - {R_q} e^{ -i \thetag \gamma_5 }   
       \end{equation}
       The phase convention for $\thetag$ is chosen such that
       $\cond{ \bar q q }$ is negative for $\thetaq=\thetag=0$.

       The generalized chiral circle theorem can be applied to induced 
       condensates of parity partners, e.g.:
       \begin{eqnarray}
          g_s \bar{q}            G_{\mu\nu} q&,& \hspace{3.5cm}
          g_s \bar{q} \gamma_{5} G_{\mu\nu} q;                        \\
          g_s \bar{q}       \sigma_{\mu\alpha} G^{\alpha}_{\nu} q 
                                - ( \mu\leftrightarrow     \nu ) &,&
                      \hspace{1cm}
          g_s \bar{q} \gamma_{5} \sigma_{\mu\alpha} G^{\alpha}_{\nu} q
                                     - ( \mu \leftrightarrow    \nu ).          
      \end{eqnarray}
      The susceptibility constants associated with these induced condensates
      are defined as follows:
      \begin{eqnarray}
          g_s \cond{ \bar{q}            G_{\mu\nu} q }  &\equiv&
        \m \m \m      \kappa_q          F_{\mu\nu} \cond{\bar q          q}
            -i \tilde{\kappa}_q \tilde{F}_{\mu\nu} \cond{\bar q \gamma_5 q}\\
          g_s \cond{ \bar{q} \gamma_{5} G_{\mu\nu} q }  &\equiv&
            -i        \xi_q     \tilde{F}_{\mu\nu} \cond{\bar q          q}
            +  \tilde{\xi}_q            F_{\mu\nu} \cond{\bar q \gamma_5 q}\\
          g_s  \bar{q}             \sigma_{\mu\alpha} G^{\alpha}_{\nu} q
                           - ( \mu\leftrightarrow\nu )  &\equiv&
        \m \m \m      \eta_q            F_{\mu\nu} \cond{\bar q          q}
            -i \tilde{\eta}_q   \tilde{F}_{\mu\nu} \cond{\bar q \gamma_5 q}
     \end{eqnarray}
      
     Using the generalized chiral circle theorem, we have
     \begin{equation}
     \tilde{\kappa}_q = \xi_q, \hspace{1cm} \tilde{\xi}_q  = \kappa_q,
                               \hspace{1cm} \tilde{\eta}_q = \eta_q.
     \end{equation}

    If we assume that all the susceptibility constants are proportional to the
    quark charge $e_q$ times a flavor independent constant, then we can rewrite
    the susceptibility constants 
    \begin{equation}
      \kappa_q = e_q \kappa, \hspace{1cm} \xi_q  = e_q \xi,
                             \hspace{1cm} \eta_q = e_q \eta.
    \end{equation}
 
    As to the second class condensates, we are only able to use the anomalous
    Ward identity to obtain the value of $\cond{G \tilde G}$, which is
    the lowest dimensional chirally invariant condensate in our
    calculation\footnote{Taking the vacuum expectation value of the anomalous
    Ward identity, we obtain $\cond{G \tilde G} \propto m_q R_q \sin
    \thetabarq$. See \cite{NEDM:chan0}.}.
    There are many unknown dimension six chirally invariant condensates, e.g.,
    $ \cond{ \bar q \gamma_\mu q \bar q \gamma^\mu \gamma_5 q}$, which cannot be
    related to their parity partners using the chiral circle theorem, and we
    shall follow the general practice to factorize these four-quark condensates
    into products of dimension three quark condensates.

    Having explained the subtlety of the calculations of NCF in QCD, we can
    present the result in terms of various coefficients of the invariant 
    tensors\footnote{We do not list the complete diagrams which contribute to 
    the OPE series of the NCFs. Instead, only terms which survive after further
    simplifications ( see discussions in the next section ) are retained in
    this section.}:

 \item {\bf Quark-Gluon Representation of the Nucleon Propagator 
            $\Pi^{(0)}_N (p)$}

       The $ f^p_1 (p^2) \m \hat{p}$ sum rule receives the contributions shown
       in Fig.1 .

       Fig.1 (a) represents the contribution of the operator $I$; 
       Fig.1 (b) represents the contribution of the operator 
                 $m_q R_q \cos \thetabarq$;
       Fig.1 (c) represents the contribution of the operator
              $R_q^2 \approx \cond{\bar q \gamma_\mu q \m \bar q \gamma^\mu q}$.

       The Feynman diagrams contribute to $ f^p_2 (p^2) e^{ i \phi^p (p^2) 
       \gamma_5 } $ sum rule are given in Fig.2.

       Fig.2 (a) represents the contribution of the operator 
                 $m_q e^{ i \thetaq \gamma_5 }$;      
       Fig.2 (b) represents the contribution of the operator
                 $R_q e^{ i \thetag \gamma_5 }$;

 \item {\bf Quark-Gluon Representation of the Polarization Tensor
            $\Pi^N_{\mu\nu} (p)$}

       Out of the six basis tensors in the polarization tensor, only three of 
       them are useful. 
       This is because the $\sigma^{\mu\nu} ( I,\gamma_5 ) $ and the
       $( p^\mu \gamma^\nu - p^\nu \gamma^\mu ) \cdot \gamma_5$ sum rules
       contain infrared singularities which indicate operator mixings with
       induced condensates whose values are unknown\footnote{For example,
       the tensor operators $m_q F_{\mu\nu}$, $m_q {\tilde F}_{\mu\nu}$ and
       $\bar q \sigma_{\mu\nu} q$, which appear in the $\sigma^{\mu\nu}
       ( I,\gamma_5 )$ sum rule, and the operators $F_{\mu\nu} G \tilde G$ and 
       $\bar q \gamma_\mu q \bar q \gamma_\nu \gamma_5 q$ in the 
       $( p^\mu \gamma^\nu - p^\nu \gamma^\mu ) \cdot \gamma_5$ sum rule, are
       mixed under renormalization group.} \cite{QSR:nmm}. 
       Consequently, we only list three sum rules below:

       The $\epsilon^{\mu \nu \alpha \beta} p_\alpha \gamma_\beta \cdot
       \gamma_5$ sum rule contains the nucleon magnetic moments and 
       the OPE diagrams are shown in Fig.3.

       Fig.3 (a) represents the contribution of the operator $F_{\mu\nu}$;
       Fig.3 (b) represents the contribution of the operator
       $m_q R_q \cos \thetabarq F_{\mu\nu}$;
       Fig.3 (c) represents the contribution of the operator $R_q^2$.

       The $\hat{p} ( p^\mu \gamma^\nu - p^\nu \gamma^\mu )$ sum rule, being
       a chirally covariant tensor, contains both the anomalous magnetic moment
       $F^N_2$ and the electric dipole moment $F^N_3$, or equivalently, 
       $F_N$ and $\alpha_N$. The diagrams in the OPE series are shown in Fig.4.

       Fig.4 (a) represents the contribution of the operator 
       $m_q e^{i\thetaq \gamma_5} F_{\mu\nu}$,
       Fig.4 (b) represents the contribution of the operator 
       $R_q e^{i\thetag \gamma_5} F_{\mu\nu}$,
       Fig.4 (c) represents the contribution of the operator 
       $\cond{\bar q \sigma_{\mu\nu} q}$.
                                                                        
\end{enumerate}

    \section{Analysis of QCD Sum Rules ( QSR ) for the 
             Nucleon Correlation Function: Part One}
    \label{sec:y5}

 The QCD sum rules ( QSR ) for the NCFs in the presence of an external 
 electromagnetic field will be analyzed in this section. 
 The main emphasis will be on
 \begin{itemize}
   \item the consistency of the QCD sum rules for NEDMs with symmetry
         constraints on strong CP violations \cite{CP:chan1}, and
   \item an order-of-magnitude estimate of hadronic variables
         ( e.g., nucleon masses $M_N$, residues $\lambda_N$, EM moments
         $\mu_N$, $d_N$ ) extracted from the QCD sum rules.
 \end{itemize}
 This is a primary analysis in the sense that we shall treat the approximate 
 sum rules ( due to truncations of the OPE series and the hadronic 
 parameterizations of the spectral functions ) as identities, and solve 
 for the hadronic observables as unknown variables in the sum rule equations.
 A more complete study will be presented in the next section.

 \begin{enumerate}
   \item {\bf QCD sum rules of the NCFs in the momentum space}

 Here we summarize the results of our calculations in the last two sections
 by listing all the invariant coefficient functions of various independent
 basis tensors\footnote{We do not list the $\sigma_{\mu\nu} ( I,\gamma_5 )$
 sum rules because of the infrared problem \cite{QSR:nmm}.} for the proton 
 correlation function in momentum space. 
 The equalities relating hadronic parameterizations and the OPE calculations for
 various invariant coefficient functions are referred to as QCD sum rules. 
 It should be kept in mind that these two representations of NCFs are derived 
 from different expansions and truncations ( a hadronic complete set in the 
 former and $1/{Q^2}$ in the latter ) of the same correlation functions. 
 Therefore, these equalities are at best approximate identities whose validity
 are empirical.

 The QCD sum rules for the proton propagator $\Pi^{(0)}_p$
 representation are given by
 \begin{equation}
  \Pi^{(0)}_p = f^p_1 (p^2) {\hat p} + f^p_2 (p^2) e^{ i \phi^p (p^2) \gamma_5 }
 \end{equation}

   (a) $ f^p_1 (p^2) $ $\mbox{   }$ ( $\hat{p}$ sum rule )

 \begin{equation}
             \frac{ \lambda_p^2   }{ p^2-M_p^2       } + \mbox{ continuum }
  = \m \frac{ p^4 \ln(-p^2) }{ 4 (2 \pi)^4     }
  +    \frac{ 4 m_u a_u \cos \thetabaru \ln(-p^2) }{ (2 \pi)^4 }
  -    \frac{ 2 a_u^2       }{ 3 (2 \pi)^4 p^2 }
 \end{equation}

   (b) $ f^p_2 (p^2) e^{ i \phi^p (p^2) \gamma_5 } $ ( $I,\gamma_5$ sum rule )

 \begin{equation}
             \frac{ \lambda_p^2  M_p   }{ p^2-M_p^2   } e^{i \theta_p \gamma_5}
           + \mbox{continuum}
  = \mbox{ } \frac{ m_d p^4 \ln(-p^2)  }{ 4 (2 \pi)^3 } e^{i \thetad  \gamma_5}
           + \frac{ a_d p^2 \ln(-p^2)  }{ (2 \pi)^4   } e^{i \thetagd \gamma_5}
 \end{equation}

 The QCD sum rules for the proton polarization tensor $\ppimunu$ in the polar 
 form representation are given by
 \begin{equation}
   \ppimunu = \sum_{i=1}^{i=6} g_i^p (p^2) \cdot T^i
 \end{equation}

   (c) $ g^p_3 (p^2) $
     ( $ \epsilon^{\mu \nu \alpha \beta} p_\alpha \gamma_\beta \cdot \gamma_5 $
         sum rule )

\begin{eqnarray}
& & \frac{ 2 \lambda_p^2 (F_1^p + F_2^p) } { 4 (p^2-M_p^2)^2 }
  + \frac{ 2 \lambda_p^2 E_A^p \cos( \frac{\varphi_A^p}{2} -
                                     \frac{\theta_p   }{2}  ) }{ 4 (p^2-M_p^2) }
  + \mbox{ continuum }               \nonumber \\
\label{eq:charge2}
&=& \m e_d  \left[ \m
      \frac{ p^2 \ln(-p^2)           }{ 2 (2 \pi)^4 }
    + \frac{ m_u a_u \cos \thetabaru }{ p^2         }
    + \frac{ a_u^2                   }{ 3 p^4       } \m \right]
    + ( \propto  e_u )         
\end{eqnarray}

   (d) $ g^p_4 (p^2) $
     ( $( p^\mu \gamma^\nu - p^\nu \gamma^\mu ) \cdot \gamma_5$ sum rule )

\begin{eqnarray}
& & \frac{ 2 \lambda_p^2 F_3^p }{ 4 (p^2-M_p^2)^2 }
  + \frac{ 2 \lambda_p^2 \mbox{ } [ E_A^p \sin( \frac {\varphi_A^p}{2} - 
                                                \frac {\theta_p   }{2}   )
                                  - E_B^p \sin( \frac {\varphi_B^p}{2} - 
                                                \frac {\theta_p   }{2}   ) ]
                                 }{ 4 (p^2-M_p^2) }
  + \mbox{ continuum }                          \nonumber \\
&=& (  \propto m_q R_q \sin \thetabarq  )
\end{eqnarray}

   (e) $ g^p_5 (p^2) + i g^p_6 (p^2) \gamma_5 $
       ( $\hat{p} ( p^\mu \gamma^\nu - p^\nu \gamma^\mu )$ sum rule )

\begin{eqnarray}
    & & \frac{ (-2) \lambda_p^2 F_p   }{ 4 M_p (p^2-M_p^2)^2 }
           e^{ i ( \theta_p + \alpha_p ) \gamma_5 }
      + \frac{   2  \lambda_p^2 E_B^p }{ 4 M_p (p^2-M_p^2)   }
           e^{ i ( \frac{\varphi_B^p + \theta_p}{2} ) \gamma_5 }
      + \mbox{ continuum }   \nonumber \\ 
    \label{eq:charge1}
    &=& e_u \left[ \frac{ m_d \ln(-p^2) }{ (2 \pi)^4 } \right] 
           e^{ i \thetad  \gamma_5 }
      +     \left[ e_d \m \frac{ \chi a_d \ln(-p^2) }{ 3 (2 \pi)^4    }
      +            e_u \m \frac{ a_d                }{ (2 \pi)^4  p^2 } 
           \right]
           e^{ i \thetagd \gamma_5 } 
\end{eqnarray}
  Here we define $a_q \equiv (2 \pi)^2 R_q \approx 0.55 ( GeV )^3$.

  The neutron sum rules can be obtained from the proton ones by doing an
  isospin rotation, namely, replacing the hadronic observables, e.g., $M_p$ by
  $M_n$, $\lambda_p$ by $\lambda_n$; and the QCD parameters, e.g., $m_d$ by 
  $m_u$,and $R_d$ by $R_u$ etc.

  \item {\bf The use of the Borel transform}

  In principle, the QCD sum rules, as summarized above, can be used to extract
  information on hadronic variables in terms of QCD parameters. However,
  the validity of the matching between these two representations for the NCFs
  is severely limited by the convergence property of the OPE series and the
  uncertainty of the higher state contributions. Fortunately, there are several
  prescriptions to improve the sum rule relations.
  Specifically, we wish to generate a set of improved sum rules such that
  the OPE series have better convergence and the contributions from a
  hadronic parametrization is dominated by the ground state observables.
  For this purpose, we use the Borel transformation, which is defined as 
  \cite{QSR:svz}:
   \begin{equation}
     {\mathcal{B}}_{M_B}[f(Q^2)]   \equiv
         \lim_{   \begin{array}{c} Q^2 \rightarrow \infty \\
                                     n \rightarrow \infty \\
                                   M_B \equiv      \frac {Q^2}{n} fixed
                  \end  {array}    }
               \left(   \frac {1}      {n!} \right) (Q^2)^{n+1}
               \left( - \frac {d}{ d{Q^2} } \right) ^{n+1}   
   \end{equation}                                                               

   We need to apply this transformation\footnote{Notice that, after the Borel
   transformation, the virtual 4--momentum variable $Q^2 \equiv -p^2$ appearing
   in the momentum space sum rules is replaced by the squared Borel mass
   $M_B^2$.} on both sides of the sum rule relations:

   For instance, on the OPE side of the QCD sum rules, we have 
 \begin{eqnarray}
 \label{eq:bope1}
    {\mathcal{B}}_{M_B}[ (Q^2)^m \ln (Q^2) ]  &=& (-1)^{m+1} m!(M_B^2)^m \\
 \label{eq:bope2}
    {\mathcal{B}}_{M_B} \left[ \left( \frac{1}{Q^2}    \right)^k \right] &=&
                         \frac{1}{(k-1)!} \left( \frac{1}{M_B^2}  \right)^k
 \end  {eqnarray}
   Thus, after Borel transformation, the higher dimensional operators in the
   OPE series receive further suppression with factorial factors.

   On the phenomenological side of the QCD sum rules, we have
 \begin{equation}
 \label{eq:bph}
    {\mathcal{B}}_{M_B} \left[ \left( \frac{1  }{Q^2+M^2} \right)^k \right]
                         =        \frac{1  }{(k-1)! }
                           \left( \frac{1  }{M_B^2  } \right)^k
                            e^{ - \frac{M^2}{M_B^2  } }
 \end{equation}
   The power suppressions of the excited state contributions are Borel
   transformed into exponential ones.

    Having established the usefulness of the Borel transformation, we need to
    specify how the matching of sum rule relations can be realized.
    As we have mentioned before, the identities we have derived cannot be
    exact for all values of $M_B^2$.
    A choice of "matching region" has to be made; such a choice is a compromise
    between different convergence properties of the two representations of the
    NCFs.
    On the OPE side, we prefer a large value of $M_B$ to suppress power
    corrections ( see Eqs.(\ref{eq:bope1}), (\ref{eq:bope2}) ); on the 
    phenomenological side, we prefer a small value of $M_B$ to enhance ground
    state observables ( see Eq.(\ref{eq:bph}) ).
    In view of this, the matching of the sum rule relations only works for a 
    finite range of the Borel mass $M_B$, and the region of matching the QCD 
    calculations and hadronic parametrizations is generally referred to as a
    Borel window.

    It is an empirical fact that within the Borel window, physical quantities
    have a mild dependence on the squared Borel mass $M_B^2$.
    Hence, in the primary analysis, we can choose a value for the squared Borel
    mass $M_B^2$ close to the ground state mass $M_N^2$ to obtained a rough
    estimation for the observables of interest.
 
    A technical remark regarding the continuum contribution:

    The continuum contributions to both nucleon propagator and the polarization
    tensor are given by the duality ansatz Eq.(\ref{eq:nuprop2}),
    (\ref{eq:polarc}). 
    After applying the Borel transform Eq.(\ref{eq:bph}) ( with $k=1$ ) on both
    sides of Eqs.(\ref{eq:nuprop2}), (\ref{eq:polarc}), we get 
     \begin{equation}
     \label{eq:borelnuprop2}
      {\mathcal{B}}_{M_B} \left[ Re \m f_i^N (p^2) ( \mbox{ Continuum } ) 
\right] \\
      =- \m \frac{1}{ \pi M_B^2 } Pr \int_{s_0^N}^\infty
    \left[ Im \m f_i^N (s) ( \mbox{ quark-gluon } ) \right]
      e^{ - \frac{s}{M_B^2} } ds
     \end{equation}

     The leading OPE terms of the NCF are of the form of $(p^2)^n \ln 
     (- p^2 - i \epsilon)$, which has an imaginary part equal to
     $- \pi (p^2)^n$. Thus the right hand side of the
     Eq.(\ref{eq:borelnuprop2})
     is a truncated Laplace transformation of a polynomial in $p^2$.
     Since the continuum contributions share a similar form as the leading OPE
     terms on the quark--gluon side of the NCF, we can combine these two terms
     together, and define the following functions:

  \begin{eqnarray}
    E_n (s,w) & \equiv & 1 - \int_w^\infty dt \m e^{-\frac{t}{s}}\m t^n     \\
    E_0 (s,w) & =      & 1 - e^{ - \frac{w}{s} }                            \\
    E_1 (s,w) & =      & 1 - e^{ - \frac{w}{s} }
                                 ( \frac{w}{s} + 1 )                        \\
    E_2 (s,w) & =      & 1 - e^{ - \frac{w}{s} } 
              ( \frac{w^2}{2s^2} + \frac{w}{s} + 1 )
\end{eqnarray}

   All functions $E_n$ act as a high--energy cutoff for the leading OPE series,
   with $E_n (s,w) \approx 0$ if $s \geq w$. Since the continuum threshold $s_0^N$ represent an average
   parameter for the excited state spectrum, we shall choose $w = s_0^N = 1.75
   GeV^2$ and write $E_n (M_B^2) \equiv E_n (s = M_B^2, w = 1.75 GeV^2)$.

   \item {\bf Borel transform--improved sum rules}

   The Borel transform--improved sum rules for the proton propagator $\ppizero$
   is given by
  
     (a) $ f^p_1 (M_B^2) $

     \begin{equation}
      M^6_B E_2 (M_B^2)
    + 4 m_u a_u \cos \thetabaru M^2_B E_0 (M_B^2) 
    + \frac{4 a_u^2}{3}
    = {\tilde \lambda}^2_p e^{ - \frac{ M_p^2 }{ M_B^2 } }
     \end {equation}

     (b) $ f^p_2 (M_B^2) e^{ i \phi^p (p^2) \gamma_5 } $

     \begin{equation}
     \label{eq:nmass}
      2 M^6_B E_2 (M_B^2)   m_d   e^{ i \thetad  \gamma_5}
    + 2 M^4_B E_1 (M_B^2)   a_u   e^{ i \thetagd  \gamma_5}
    = {\tilde \lambda}^2_p  M_p   e^{ - \frac{ M_p^2 }{ M_B^2 } }
                                e^{ i \theta_p \gamma_5}
     \end {equation}

     where ${\tilde \lambda}^2_N \equiv 2 (2 \pi)^4 \lambda^2_N$.
     For the proton polarization tensor $\ppimunu$, we obtain the following sum
     rules:

     (c) $ g^p_3 (M_B^2) $

      \begin{eqnarray}
     & &  e_u M_B^4 E_1 (M_B^2) + e_d m_u a_u \cos \thetabard +
         \frac{ a_u^2 }{ 3 M_B^2 } [ - ( e_d + \frac{2 e_u}{3}) + 
                     ( \propto e_u ) ]
         \nonumber \\
      \label{eq:nmm}
     &=& \frac{ {\tilde \lambda}^2_p }{4} e^{ - \frac{ M_p^2 }{ M_B^2 } }
       \left[ \m \frac{ F_1^p + F_2^p }{M_B^2}
       + E_p^A \cos( \frac{\varphi_p^A}{2} - \frac{\theta_p}{2} ) \m \right] 
      \end  {eqnarray}

     (e) $ g^p_5 (p^2) + i g^p_6 (p^2) \gamma_5 $

      \begin{eqnarray}
      \label{eq:npt}
     & & \left[ \m 4 e_u a_d + ( \propto e_d ) \m \right]
          e^{ i \thetagd  \gamma_5}
      -  \left[ \m 4 e_u m_d + ( \propto e_d ) \m \right] M_B^2
          e^{ i \thetad   \gamma_5} 
        \nonumber     \\
     &=& \frac{ {\tilde \lambda}^2_p }{ M_p } e^{ - \frac{ M_N^2 }{ M_B^2 } }
         \left[ \m \frac{F_p}{M_B^2}
                e^{ i (           \theta_p + \alpha_p     ) \gamma_5 } 
                        +  E_p^B 
                e^{ i ( \frac {\varphi_p^B + \theta_p}{2} ) \gamma_5 } \m
         \right] 
      \end  {eqnarray}

  \item{\bf Manipulation of the QCD Sum Rules}

    While we have identified the hadronic unknowns and written down the sum
    rule equations to relate hadronic observables in terms of quark--gluon
    parameters, we need to take some further simplifications to make the
    solutions of these complicated relations manageable\footnote{However, we
    emphasize that none of these simplifications are absolutely necessary for
    the physical content of the further discussions. The choice of these
    simplifications should be regarded as pedagogic and to help us isolate
    issues of strong CP violation from other complications in the strong
    interactions.}.

    The three steps we adopt to simplify the sum rule relations are discussed
    in order:

    \begin{enumerate}

    \item The use of isospin symmetry

          It is important that the symmetry constraints on the strong CP
          violations require that all quarks have finite masses.
          In particular, the existence of any massless quark leads to no strong
          CP violation. 
          In such a limit, the flavor symmetry is broken and we have to make
          the flavor dependence in our calculation explicit.
          While it is possible to realize these limits in the QCD sum rule
          approach \cite{NEDM:chan0}, we opt for isospin symmetry to simplify
          our calculations.
          Consequently, for the quark-gluon parameters, we take 
          $m_u = m_d = m_q$ and $R_u = R_d = R_q$.
          It can be shown that in this case, $\thetau$ can be chosen to be
          equal to $\thetad$ and, as a consequence of Crewther's condition,
          $\thetagu = \thetagd = \thetag/2$. 
          See Eqs.(\ref{eq:crewther1}) (\ref{eq:crewther2}).

          On the other hand, since we do not include QED corrections, which
          break the isospin symmetry, in the calculations of the hadronic
          masses and residues, we can take $M_p = M_n$ and
          $\lambda_p = \lambda_n$.
          Thus, the use of isospin symmetry greatly reduces the unknowns and
          the input parameters in the sum rule relations.

          The other advantage of this simplification, as pointed out by 
          B.L. Ioffe and A.V. Smilga \cite{QSR:nmm}, is that we can eliminate
          the induced condensates which generate the unknown susceptibility
          constants from the proton and neutron sum rules by taking certain
          combinations with different quark charge dependences.

          For example, in the 
          $\hat{p} ( p^\mu \gamma^\nu - p^\nu \gamma^\mu ) ( 1, \gamma_5 )$ 
          sum rule we can multiply the proton sum rule Eq.(\ref{eq:charge1}) by
          $e_u$ and subtract the corresponding neutron sum rule multiplied by
          $e_d$. 
          This eliminates the contributions of the induced condensates
          $\cond{\bar q \sigma_{\mu\nu} q} \propto \chi \cond{\bar q q}$.
          For the 
          $\epsilon^{\mu \nu \alpha \beta} p_\alpha \gamma_\beta \cdot \gamma_5$
          sum rule Eq.(\ref{eq:charge2}), we repeat the similar procedure, but
          with $e_d$ times the proton sum rule minus $e_u$ times the neutron
          one.
          As a bonus, we also eliminate the infrared singularity coming
          from the operator $m_q F_{\mu\nu}$.

    \item The elimination of excited state parameters in the polarization tensor

          We shall not concern ourselves with the details of the excited states
          in this calculation.
          With the assumption that the ground state to excited
          state transition can be approximated by momentum independent
          constants Eq.\ref{eq:excited constant}, one can observe that such a
          contribution has a different Borel mass dependence, as compared to the
          ground state observables, see Eqs.(\ref{eq:nmm}) (\ref{eq:npt}). 
          We can apply the differential operator
          \begin{equation}
            1 - M_B^2 \frac{\partial}{\partial M_B^2}
          \end{equation}
          on both sides of the sum rule equations Eqs.(\ref{eq:nmm}), 
          (\ref{eq:npt}) to eliminate the contributions of ground to excited 
          state transitions in the sum rule relations.    
 
    \item The problem of operator mixing in the
          $( p^\mu \gamma^\nu - p^\nu \gamma^\mu ) \cdot \gamma_5$ sum rule

          After all these simplifications, we have seven unknowns
          $\lambda_N,$$M_N,$$\theta_N,$$F_p,$$\alpha_p,$$F_n,$$\alpha_n$
          in the sum rules relations.
          While the nucleon propagator $\Pi^{(0)}_N$ gives three independent
          sum rules ( $\hat p, 1, \gamma_5$ ) for the nucleon variables 
          $\lambda_N, M_N, \theta_N$, we have
          only three useful identities Eqs.(\ref{eq:nmm}),(\ref{eq:npt}) in the
          polarization tensor $\Pi_{\mu\nu}^N$ for four EM moments 
          $F_p, \alpha_p, F_n, \alpha_n$!
          The leading OPE contribution to the tensor
          $( p^\mu \gamma^\nu - p^\nu \gamma^\mu ) \cdot \gamma_5$ is 
          proportional to the operator $F_{\mu\nu} \cond{G \tilde G}$, whose 
          Wilson coefficient has a $\ln m_q$ infrared singularity\footnote{We
          need to calculate a quark propagator in the presence of two or three
          constant external fields ( $G_{\mu\nu}$ and/or $F_{\mu\nu}$ ), which
          is infrared divergent if the quark mass is zero.}. 
          One can argue that such an infrared singularity should be included in
          the definition of QCD condensates, which requires a separation scale
          instead of a small quark mass.
          The difficulty here is to show our calculation is independent of the
          choice of a separation scale, as required by the renormalization
          group analysis.
          The situation is further complicated by the mixing of operators under
          renormalization.
          Without a detailed analysis of the operator mixing problem, we
          can not obtain useful information from this tensor.
          Consequently, we need to introduce a phenomenological parameter $\beta$ for
          the ratio of $F^p_3$ to $F^n_3$.
          By comparing the Feynman diagrams contributing to the tensor 
          $( p^\mu \gamma^\nu - p^\nu \gamma^\mu ) \cdot \gamma_5$ for both proton
          and neutron, we find the only difference is the charge dependence of u and 
          d quarks, the quark masses and chiral radii do not enter this sum rule
          because of the chiral property of this sum rule. 
          The charge dependence seems to indicate that $\beta = {e_u}/{e_d}$.
          Henceforth, we shall assume such a relation in our subsequent discussion.
           
    \end{enumerate}

  \item{\bf Symmetry Constraints from the QCD Sum Rule Relations}

    Despite the complicated appearance of the sum rule relations, it is
    possible to show that, due to the use of the polar form in both hadronic and
    quark--gluon representations of the NCF, our sum rule calculations
    satisfy the symmetry constraints on the strong CP violation \cite{CP:chan1}
    without an explicit solution of the CP violating observable, e.g.,NEDM.

    To see this, we focus on the $(1, \m \gamma_5)$ sum rules, 
    Eq.(\ref{eq:nmass}) in the nucleon propagator and the 
    $\hat{p} ( p^\mu \gamma^\nu - p^\nu \gamma^\mu ) (1, \m \gamma_5)$ sum rules,
    Eq.(\ref{eq:npt}) in the polarization tensor. 
    Both sets of sum rules are associated with chirally covariant tensors and receive
    contributions from chirally covariant condensates.
    In a shorthand notation, we can rewrite the sum rules in the following forms:
 \begin{eqnarray}
 \label{eq:ab eq}
      A  \m m_q \m e^{i \thetaq \gamma_5} + B  \m R_q \m e^{i \thetagq \gamma_5}
  &=& C  e^{i   \theta_N              \gamma_5}    \\
 \label{eq:a'b' eq}
     A' \m m_q \m e^{i \thetaq \gamma_5} + B' \m R_q \m e^{i \thetagq \gamma_5}
  &=& C' e^{i ( \theta_N + \alpha_N ) \gamma_5}    
 \end{eqnarray}
    We emphasize that the structure of the sum rules, as summarized above, is
    independent of the approximation scheme which only affects the values of 
    the numerical coefficients $A, B, A'$ and $B'$. Specifically, in terms of
    the Wilson coefficients appearing in the ( Borel transformed ) QCD sum 
    rules, these numerical coefficients are
  \begin{eqnarray}
    \label{eq:num1}
    A &=&  2 M_B^6 E_2(M_B^2)                       \\
    \label{eq:num2}
    B &=&  2 M_B^4 E_1(M_B^2)                       \\
    \label{eq:num3}
    A'&=&  4 ( e_d^2 - e_u^2 ) M_B^4 E_0(M_B^2)     \\
    \label{eq:num4}
    B'&=& -4 ( e_d^2 - e_u^2 ) M_B^2                
  \end{eqnarray}

    In the two sets of sum rules Eqs.(\ref{eq:ab eq}) (\ref{eq:a'b' eq}), we have 
    four hadronic unknowns $C, C', \theta_N,$ and $\alpha_N$ to be determined.
    The relation with the nucleon variables is that $C$ is a function of $\lambda_N$
    and $M_N$, $C'$ is a function of $F_p$ and $F_n$, with an 
    assumption\footnote{This assumption is chosen for its simplicity and is not
    equivalent to the conjectured value for $\beta =  {e_u}/{e_d}$.} that
    \begin{equation}
    {\tan}^{-1} \frac{F^3_p}{F^2_p} = {\tan}^{-1} \frac{F^3_n}{F^2_n} \equiv \alpha_N
    \end{equation}
    we rearrange the seven nucleon variables into four unknowns.

    The former sum rule, Eq.(\ref{eq:ab eq}) contains the nucleon chiral phase
    $\theta_N$, the latter Eq.(\ref{eq:a'b' eq}) has both $\theta_N$ and $\alpha_N$,
    in the hadronic representations of the NCF.
    It is the relative ( chiral ) phase difference that defines the physically
    measurable ( and chirally invariant ) dimensionless ratio ${d_n}/{\mu_n^a}$,
    which determines the violation of the CP symmetry.

    On the other hand, the OPE series for both sum rules are organized in such 
    a way that the explicit chiral symmetry breaking parameter $m_q$ and the 
    spontaneous chiral symmetry breaking parameter $R_q$ are on equal 
    footings\footnote{The form of the chirally covariant sum rules is a concrete
    realization of the strong CP torus of QCD, as described in refer. 
    \cite{CP:chan1}.}. 
    Such a representation of the NCF is particularly useful
    for examining the symmetry constraints on the strong CP violation.
    Namely, in either limit $m_q \rightarrow 0$ or $R_q \rightarrow 0$,
    the nucleon chiral phase $\theta_N$ becomes $\thetagq$ or $\thetaq$,
    respectively, and the relative phase $\alpha_N \equiv {\tan}^{-1} 
    ({d_n}/{\mu_n^a})$ has to vanish. 
    Thus, there is no strong CP violation if chiral symmetry is exact in QCD
    and CP violating observables must be proportional to the product of $m_q$
    and $R_q$. 
    Furthermore, the periodic structure in the chiral phases naturally
    generates a $\sin\thetabar$ factor\footnote{This is the simplest periodic
    function, which is regular and approaches zero as $\thetabar$ vanishes.},
    which combined with the two chiral symmetry breaking parameters $m_q$ and 
    $R_q$, implies that CP violating observables are proportional to 
    $\langle G \tilde G \rangle_{\thetabar}$\footnote{Taking the expectation
    value of the anomalous Ward identity for the flavor singlet axial current,
    one can show that  $\langle G \tilde G \rangle_{\thetabar}$ is propotional
    to the product of $m_q$, $R_q$ and $\sin\thetabar$.}.

    We can solve the equations (\ref{eq:ab eq}), (\ref{eq:a'b' eq}) for the
    hadronic unknowns $C, C', \theta_N, \alpha_N$ as follows:
  \begin{eqnarray}
   C_N^2 &=&    A^2 m_q^2 +    B^2 R_q^2 + 2 A  B  m_q R_q \cos \thetabar  \\
{C'}_N^2 &=& {A'}^2 m_q^2 + {B'}^2 R_q^2 + 2 A' B' m_q R_q \cos \thetabar  \\
  \tan \theta_N &=& \frac{A m_q \sin \thetaq + B R_q \sin \thetag}
                         {A m_q \cos \thetaq + B R_q \cos \thetag}         \\
  \label{eq:sina}
  \tan \alpha_N &=& \frac{ ( A B' - B A' ) m_q R_q \sin \thetabar }
                           { A A' m_q^2 + B B' R_q^2 +
                           ( A B' + B A' ) m_q R_q \cos \thetabar } 
    \end{eqnarray}

   The nucleon variables are given by the following expressions:
   \begin{eqnarray}
   \label{eq:Borel function1}
   M_N &=& \frac{ C_N }{ M^6 E_2 (M_B^2) + 4 m_q R_q M_B^2 E_0 (M_B^2) 
           \cos \thetabarq + 4 R_q^2 / 3 }      \\
 \lambda_N &=& M_B^6 E_2 (M_B^2) + 4  m_q R_q M_B^2 \cos \thetabarq E0(M_B^2)
                + 4 R_q^2 / 3                  \\
  F_2^p &=& \frac{M_N^2}{ e_u^2 - e_d^2 } ( e_u U(M_B^2, \thetabarq) -
                                            e_d K(M_B^2, \thetabarq) + e_d^2)\\
  F_2^n &=& \frac{M_N^2}{ e_u^2 - e_d^2 } ( e_d U(M_B^2, \thetabarq) -
                                            e_u K(M_B^2, \thetabarq) 
                                          + e_d e_u)\\
  F_3^p &=& \frac{M_N^2}{ e_u^2 - e_d^2 } e_u V(M_B^2, \thetabarq) \\
  \label{eq:Borel function2}
  F_3^n &=& \frac{M_N^2}{ e_u^2 - e_d^2 } e_d V(M_B^2, \thetabarq)
   \end{eqnarray}
 where
  \begin{eqnarray}
     K(M_B^2, \thetabarq) &\equiv& 
      \left( 1 - M_B^2 \frac{\partial}{\partial M_B^2} \right)
      (e_u^2 - e_d^2) \left[
      \frac{4 R_q^2 / 3 + 4 m_q R_q M_B^2 E_0 (M_B^2) \cos \thetabarq }
           { M^6 E_2 (M_B^2) + 4 m_q R_q M_B^2 E_0 (M_B^2) \cos \thetabarq 
              + 4 R_q^2 / 3 }  \right]\\
     U(M_B^2, \thetabarq) &\equiv&
      \left( 1 - M_B^2 \frac{\partial}{\partial M_B^2} \right)
      \left[ \frac{A A' m_q^2 + ( A B' + A' B ) m_q R_q \cos \thetabarq +
                   B B'  R_q^2}{C_N^2} \right]    \\
     V(M_B^2, \thetabarq) &\equiv&
      \left( 1 - M_B^2 \frac{\partial}{\partial M_B^2} \right)
      \left[ \frac{ ( A B'- A' B ) \cond{G \tilde G} }{ 2 C_N^2 } \right]
  \end{eqnarray}

    Given the small value of the current quark mass, $m_q \approx 5 MeV$, and
    the relative minus sign in the numerator of Eq.(\ref{eq:sina}),
    it is possible that the CP violating observables could be dynamically
    suppressed without a tiny $\thetabar$ parameter.
    We shall examine such an interesting scenario in the next section.

  \item{\bf Order of Magnitude Estimations of the Hadronic Observables from
            the QCD Sum Rules}

    Before we embark on a complete analysis of the sum rule relations, it is
    useful to estimate the relative sizes of the hadronic observables. 
    The qualitative features of the estimations also provide a test whether the
    underlying assumptions in the sum rule calculation work for the case in 
    which we are interested.
    It is an empirical fact that the Borel window lies around the $1 GeV$ 
    region and we can choose a value for the Borel mass $M_B$ to be the nucleon
    mass $\approx 1 GeV$, if the Borel mass dependence of the hadronic
    observables is weak. 
    Other inputs needed for the estimations of the hadronic observables include
    the current quark mass $m_q \approx 5 MeV$, and the quark chiral radius
    $R_q \approx - \langle \bar q q \rangle_{ \thetaq = \thetag = 0 } =
    (240 MeV)^3$. The latter gives $a_q = 0.55 (GeV)^3$.

    Substituting these numbers into Eqs.(\ref{eq:num1}) $\sim$ (\ref{eq:num2}),
    in the small $\thetabar$ limit, the functions 
  Eqs.(\ref{eq:Borel function1}) $\sim$ (\ref{eq:Borel function2}) reduce to
  \begin{eqnarray}
       M_N (\thetabar) &=&   0.9  GeV    \\
   {\tilde \lambda}^2 (\thetabar) &=&  2.5  (GeV)^6 \\
     F_2^p (\thetabar) &=&       3.1     \\
     F_2^n (\thetabar) &=&     - 2.1       \\
     F_3^p (\thetabar) &=&             10^{-2}  \sin \thetabar     \\
     F_3^n (\thetabar) &=&     - 0.5 \times  10^{-2}  \sin \thetabar     
  \end{eqnarray}

   Comparing the ratio ${F_3^N}/{F_2^N}$ with the current experiment upper
   bound on the nEDM ${d_n}/{\mu_n^a} \leq  10^{-11}$, we obtain an upper
   bound on the strong CP violating parameter $\thetabar \leq 10^{-9}$.

   We emphasize that the upper bound on the $\thetabar$ parameter is obtained from a
   calculation without assuming a perturbative expansion on the $\thetabar$
   parameter.
   Consequently, we can derive an upper bound on the ratio ${d_n}/{\mu_n^a}$ with
   respect to the $\thetabar$ parameter. Such information is an intrinsic property of
   the CP conserving ( $\thetabar=0$ ) QCD and provides a dynamical suppression 
   mechanism to the CP violating observables for solving the strong CP problem.
   Unfortunately, the number we have derived
   $\max_{0 \leq \thetabar \leq 2 \pi} \m \frac{d_n}{\mu_n^a} \m \leq 10^{-2}$, 
   is not small enough to achieve this end.
   Therefore, we conclude that the solution to the strong CP problem has to lie
   beyond QCD. 

\end{enumerate}

    \section{Analysis of QCD Sum Rules ( QSR ) for the 
             Nucleon Correlation Function: Part Two}
    \label{sec:y6}

 This section aims at a more careful study of the QSRs for the NCFs in the
 presence of an external EM field. 
 We shall extract various nucleon variables,
 e.g., the nucleon masses $M_N$, nucleon residues $\lambda_N$ and their
 EM moments $F_2^N, F_3^N$, from the QSRs in a more rigorous manner.
 The emphasis here is to use the QCD sum rule method to obtain quantitive 
 results for the hadronic variables, including an analysis and/or estimate of
 the errors and uncertainties in our calculations.

 We first discuss how to choose a region for the matching between two 
 representations of the NCFs; then we extract the hadronic observables from a 
 least square fit.
 The possible errors and uncertainty are discussed in the third part and we 
 summarize the results obtained from the sum rule analysis in the last part.
\begin{enumerate}

 \item {\bf Choice of the Borel Window}

 As we have emphasized before, the hadronic and quark-gluon representations of
 the NCFs are based on two different expansion schemes and there is no a priori
 reason that, under our truncation and approximation, these two expansions
 should provide the same information of the NCFs. 
 While it is possible to enlarge the overlap and improve the convergence with 
 the use of the Borel transform, a matching region between both sides of the
 sum rules has to be specified. 
 In view of the different convergence properties, we use the following
 criterion to define the Borel window.

 First of all, on the quark gluon representation of the NCFs, we require that
 the highest dimension condensates contribute to the whole series less than 
 $20\%$.
 This leads to a lower bound on the Borel mass squared $M_B^2 \geq \m 
 \mbox{0.8 GeV}^2$.

 Secondly, for the hadronic parametrization, we restrict the contribution from 
 the continuum to be no greater than $20\%$ of the leading OPE contributions to
 the NCF.
 This leads to an upper bound on the Borel mass squared $M_B^2 \leq \m 
 \mbox{1.2 GeV}^2$.

 Thirdly, the extractions\footnote{There is no standard procedure to extract
 hadronic variables in the sum rules approach. In our case, we average the
 hadronic variables over the Borel window and use the $\chi^2$ to represent the
 uncertainity with our extractions.} of the hadronic variables in the region 
 specified above introduce additional uncertainity. 
 Presumably, the third uncertainity is not independent from the ones associated 
 with the convergence requirement. 
 However, such a correlation is not transparent in our analysis, and we simply
 take the intersection region of the three criterions to define the Borel
 window, which lies between $0.8 {GeV}^2$ to $1.2 {GeV}^2$.

\item {\bf Extraction of the Hadronic Observables from the QCD Sum Rules}

 Given the simplifications discussed in the previous section, we obtain several
 relations which give the hadronic observables as
 functions of the squared Borel mass.
 As we have emphasized, if the matching between the hadronic and quark--gluon 
 representations of the NCF is realized in the given Borel window, all the 
 hadronic observables should have only a mild dependence on the squared Borel
 mass. 
 Therefore, we plot\footnote{In presenting these figures, we have chosen the
 value of the $\thetabar$ parameter to be $10^{-10}$.} all six chirally
 invariant hadronic observables, nucleon mass $M_N$, nucleon residue 
 ${\tilde\lambda}_N^2$, proton and neutron magnetic moments $1+F_2^p, F_2^n$ 
 and electric dipole moments $F_3^p, F_3^n$, as functions of the squared Borel
 mass $M_B^2$ in Fig.5 and Fig.6. 
 The final values of these hadronic observables are determined by averaging 
 the functions over the Borel window and the uncertainty is given by the 
 $\chi^2$ of this average.
 Since it is necessary to have such $\chi^2$ uncertainty to be smaller than the 
 uncertainty associated with the choice of Borel window, we shall consider the
 result obtained from our sum rule analysis to lie within $20\%$ uncertainty.
 
\item {\bf Error Analysis}

     We list a few comments on the errors and uncertainties related to this
     calculation.

     On the OPE calculations of the NCF:

       (1) The convergence of the OPE expansion:

           The OPE series for the NCF in momentum space is a 
           ${\Lambda_{QCD}^2}/{Q^2}$ expansion, where we assume that all vacuum
           condensates are proportional to $\Lambda_{QCD}$ to a certain
           power, with a numerical coefficient of order $O(1)$.
           While there is no systematic way to estimate the large order
           behavior of the coefficients and test this assumption,
           the expansion series could be an asymptotic series and has no
           convergence radius at all.
           We have to content ourselves with the naive estimate from the Borel
           window analysis and estimate the contributions from the higher 
           dimensional condensates to be below $20\%$.

       (2) Approximations in the calculations of the Wilson coefficients:

          When we calculate the OPE series of a correlation function, the Wilson
          coefficients are expanded in the powers of "small" parameters,
          $\alpha_s$, $m_q^2 / {Q^2}$ and we make truncations of the series to
          obtain an approximate results for the Wilson coefficients. 
          Again, the higher order behavior of the perturbative expansion is
          poorly known and we make no attempt to study the higher order 
          contributions in the present work.

       (3) Hadronic parameterizations:

          This is the most challenging part in the sum rule analysis, because we
          have no ad hoc criterion to see how well our ansatz, e.g., the
          momentum independence of the excited state parameters $E^N_A, E^N_B$,
          works.
          There are some analyses which show that this part could bring in a 
          large uncertainty to a sum rule calculation \cite{QSR:excitation}.
     
       (4) The extraction of physical quantities from the matching:
  
         Since physical quantities should be independent of the Borel mass,
         the deviation of the physical observables from a constant value in a
         given Borel window should not be larger than the uncertainty associated
         with the choice of a Borel window. 
         Nevertheless, the scheme dependent errors ( how do we extract
         physical quantities from the matching ) are related to the $\chi^2$ of
         the matching, and represent the quality of a sum rule calculation.

       (5) The dependence on the input parameters:

         The QCD sum rule calculations generally rely on input parameters which
         are derived from other sources, e.g., the quark mass, quark condensates
         etc.. 
         In many cases, these numbers are not well determined and their
         variations from a "standard" value have to be taken into account when
         we estimate the errors of the physical quantities from a QCD sum
         rule calculation.
         Recently, an analysis was performed by D.B. Leinweber \cite{QSR:DBL} 
         and we shall not dwell on this aspect further in this work.
      
\item{\bf Summary of the Results}

        In summary, we obtain the following results for the hadronic
        observables from the sum rule analysis:
        \begin{eqnarray}
         M_N &=&  0.9     GeV   \\
         {\tilde \lambda_N}^2 &=& 2.5GeV^6  \\
         F^p_2 &=& 3.1     \hspace{3.5cm} F^n_2 = -2.1\\
         F^p_3 &=& 10^{-2} \thetabar     \hspace{2cm} F^n_3 = -0.5 \times 
10^{-2} \thetabar
         \end{eqnarray}

\end{enumerate}

    \section{Summary and Conclusion}
    \label{sec:y7}


The electric dipole moments of nucleons ( NEDM, $d_N$ ) are calculated using 
the method of QCD sum rules. 
Through the use of a polar representation of both nucleon EM moments 
( anomalous magnetic moment $F_2^N$ and electric dipole moment $F_3^N$ ) and
the $U_A(1)$ covariant quark condensates, we are able to demonstrate the
reparameterization invariance of CP violating observables explicitly in a QCD
Lagrangian with a CP violating $\thetabar$ parameter.
The symmetry constraints on strong CP violation in QCD, together with a dual
relationship between the quark mass $m_q$ and the chiral radius $R_q$ 
\cite{CP:chan1} are realized in the QCD sum rule relations which connect the 
hadronic observables to the QCD parameters.
The extraction of the NEDM in terms of the $\thetabar$ parameter and QCD 
parameters from the QCD sum rule relations can be achieved without assuming a 
perturbative expansion of the $\thetabar$ parameter.
In addition, the periodic dependence of the CP violating observables on the 
$\thetabar$ parameter comes out naturally in our approach.

Our final result establishes a functional dependence of the NEDM on the 
$\thetabar$ parameter, which, combined with the experimental upper bound on the
nEDM, gives an upper bound on the unknown $\thetabar$ parameter of less than
$10^{-9}$. 
While this result is compatible with previous calculations of the nEDM using 
different techniques, it also indicates that a dynamical suppression of the CP 
violating observables, as implied by the symmetry constraints on the strong CP
violation in QCD, is not sufficient to resolve the strong CP problem.
Therefore, we conclude that a solution to the strong CP problem does not exist
within QCD and a natural explanation of the tiny $\thetabar$ parameter 
necessarily requires physics beyond the standard model.
     
    \acknowledgments

C-T Chan thanks Prof. T. Hatsuda, Prof. W.C. Haxton, and 
Prof. E.G. Adelberger, for many useful comments and discussions.
This work was supported by Nuclear Theory Group of Department of Physics at
University of Washington, under the grant DE-FG-03-97ER41014.

    \begin{figure}
      \caption{Feynman Diagrams which contribute to the $\hat p$ sum rule}
    \end{figure}

    \begin{figure}
      \caption{Feynman Diagrams which contribute to the $1$ and $\gamma_5$ 
               sum rules} 
    \end{figure}

    \begin{figure}
      \caption{Feynman Diagrams which contribute to the
               $\epsilon^{\mu\nu\alpha\beta} p_\alpha \gamma_\beta \gamma_5$
               sum rule}
    \end{figure}

    \begin{figure}
      \caption{Feynman Diagrams which contribute to the
          $\hat p ( p^\mu \gamma^\nu - p^\nu \gamma^\mu ) ( 1,\m \gamma_5 )$ 
               sum rule}
    \end{figure}

    \begin{figure}
      \caption{The nucleon mass $M_N$ as a function of the squared Borel mass
               $M_B^2$}
    \end{figure}
    
    \begin{figure}
      \caption{The nucleon residue ${\tilde \lambda}_N^2$ as a function of
               the squared Borel mass $M_B^2$}
    \end{figure}

    \begin{figure}
      \caption{The nucleon magnetic moment $F_1^N + F_2^N$ as a function of the
               squared Borel mass $M_B^2$}
    \end{figure}

    \begin{figure}
      \caption{The nucleon electric dipole moment $F_3^N$ as a function of the
               squared Borel mass $M_B^2$}
    \end{figure}

    \begin{table}
      \caption{Nucleon to higher state EM transition matrix element}
    \end{table}

\newpage
  
   \begin{figure}

     \centerline{\psfig{file=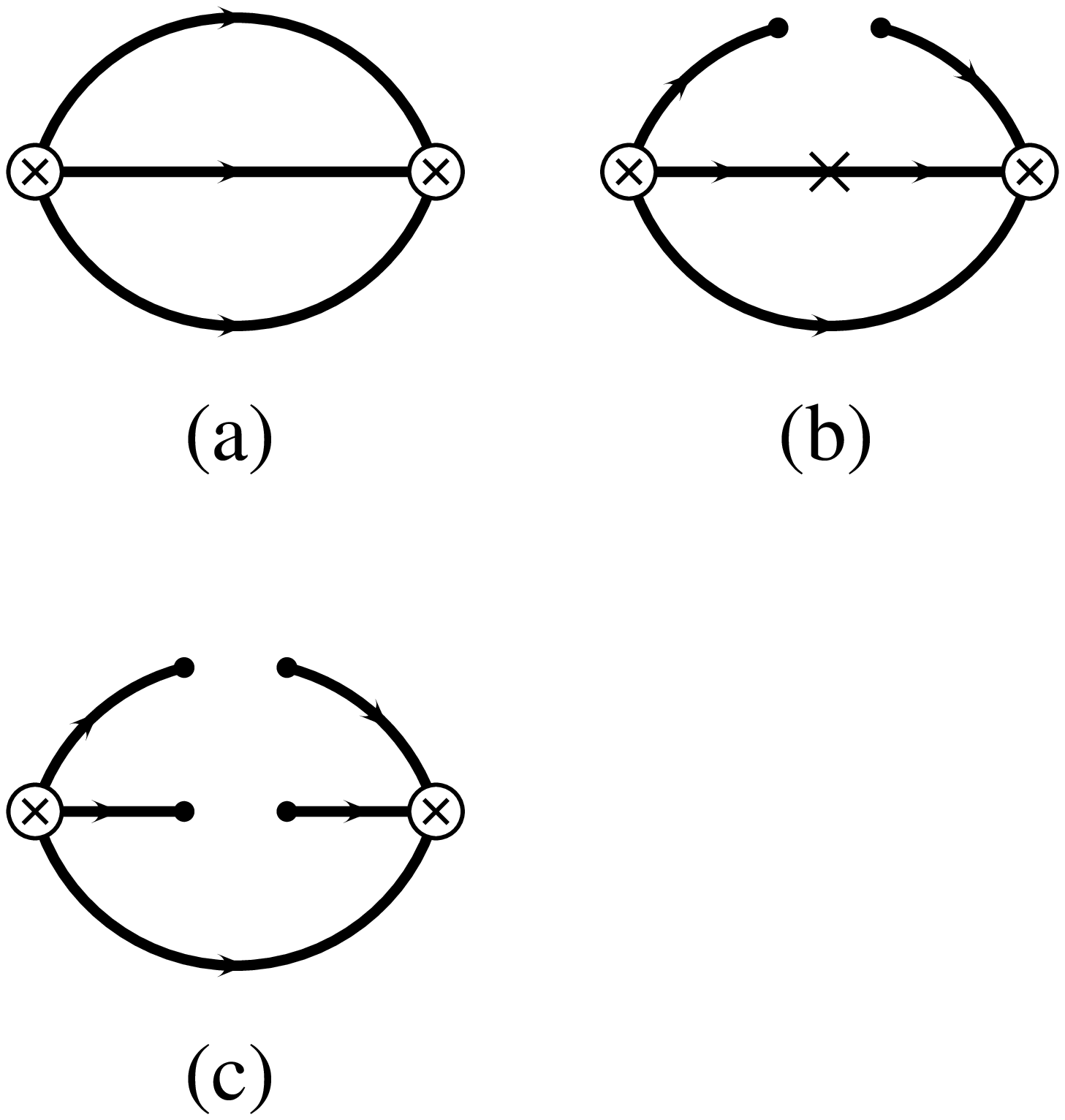}}

   \end{figure}

   FIG.1

\newpage

   \begin{figure}

     \centerline{\psfig{file=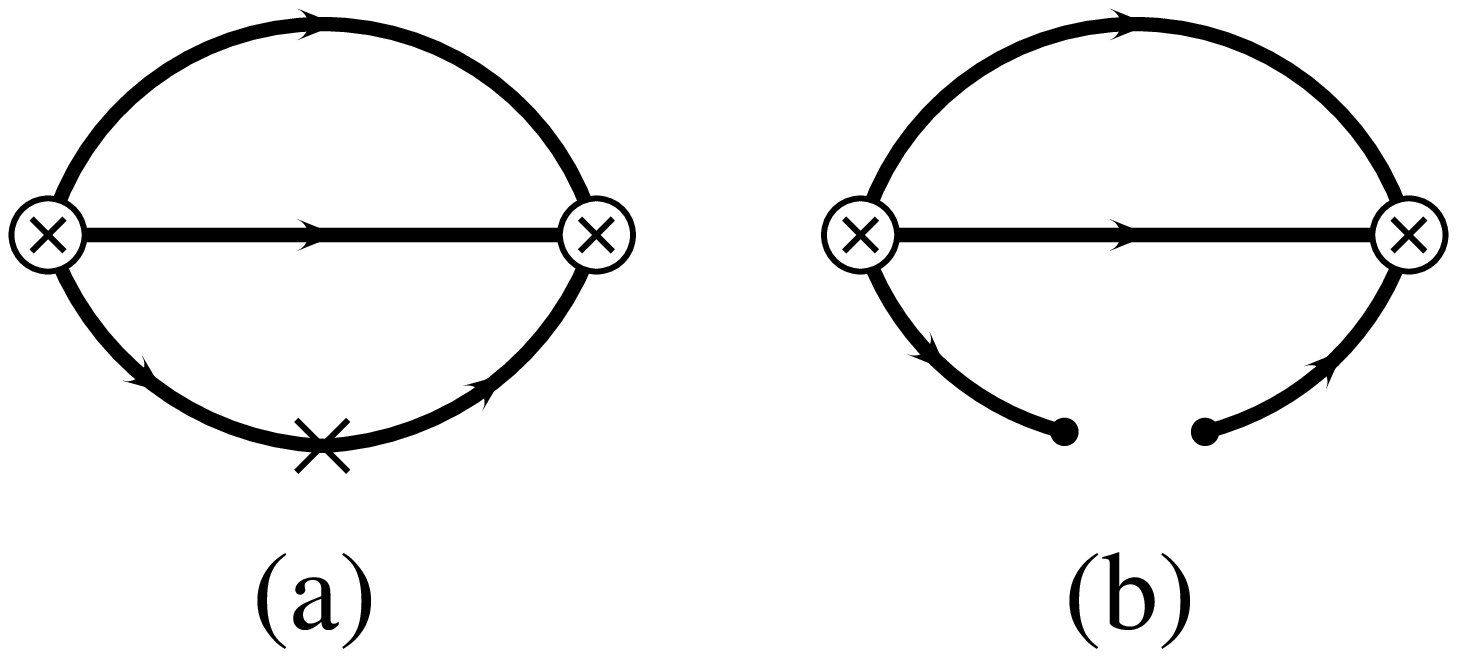}}
    
    \end{figure}

   FIG.2

\newpage

   \begin{figure}

     \centerline{\psfig{file=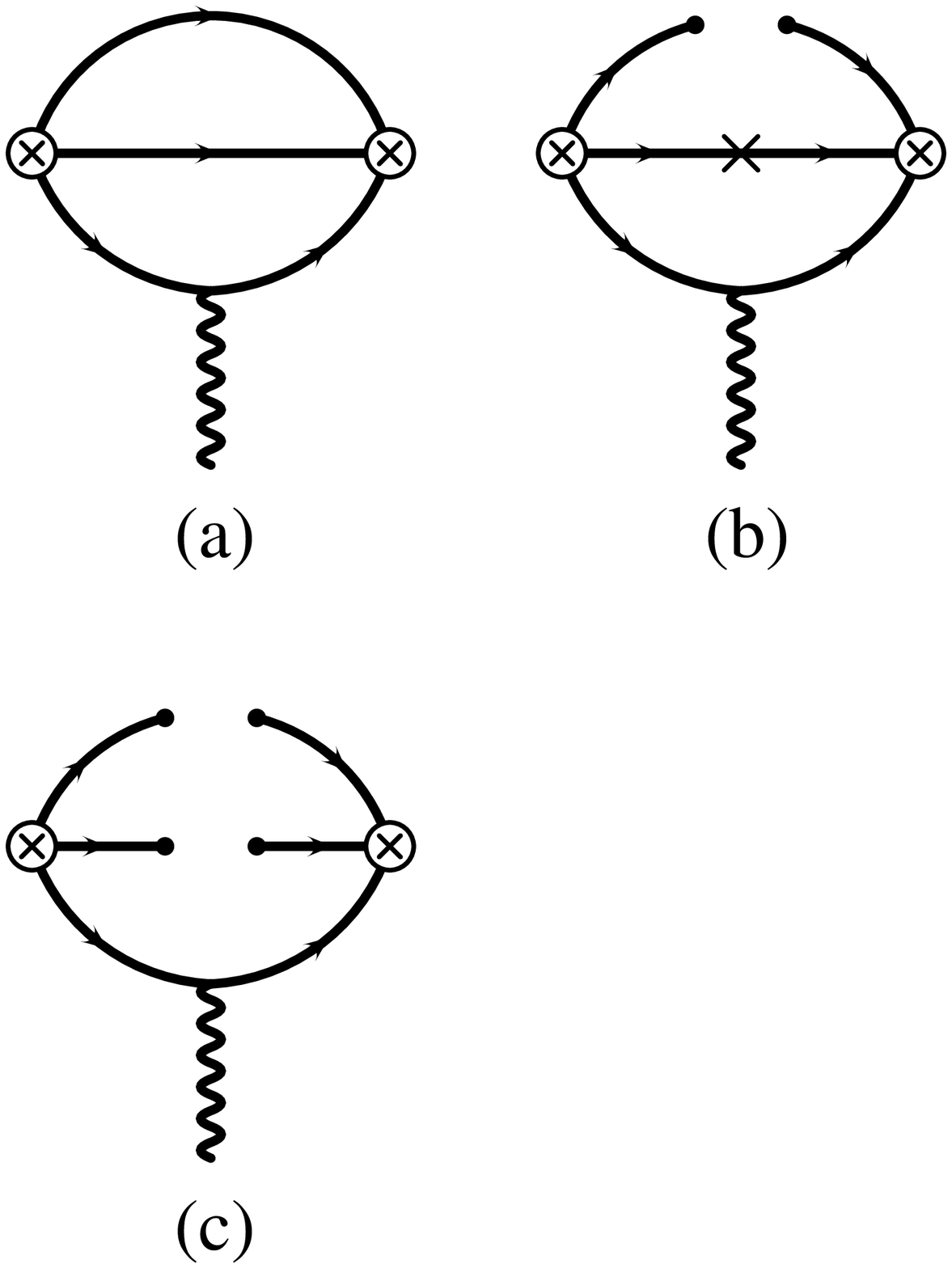}}

   \end{figure}

   FIG.3

\newpage

   \begin{figure}

     \centerline{\psfig{file=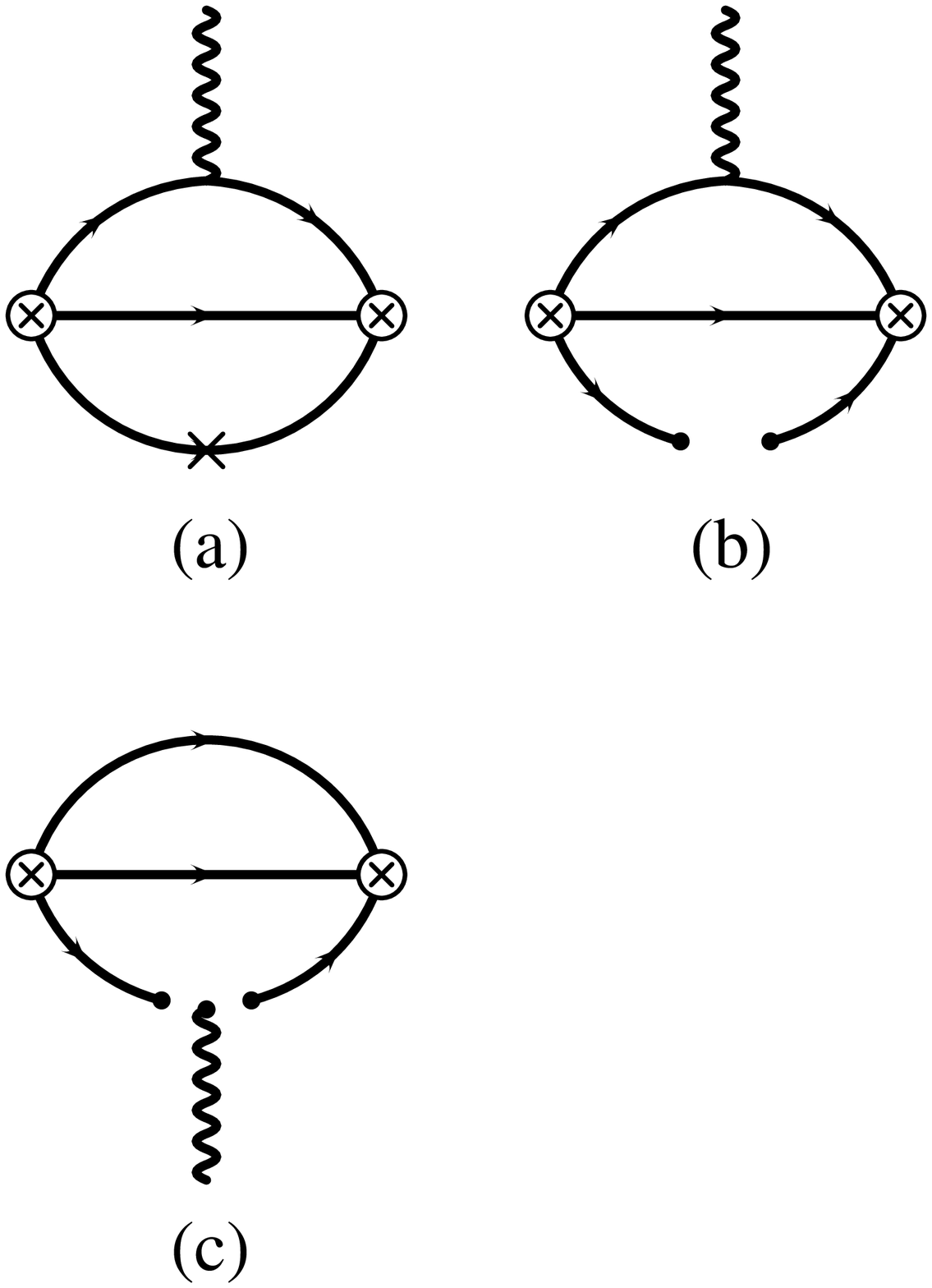}}

   \end{figure}

   FIG.4

\newpage

   \begin{figure}

     \centerline{\psfig{file=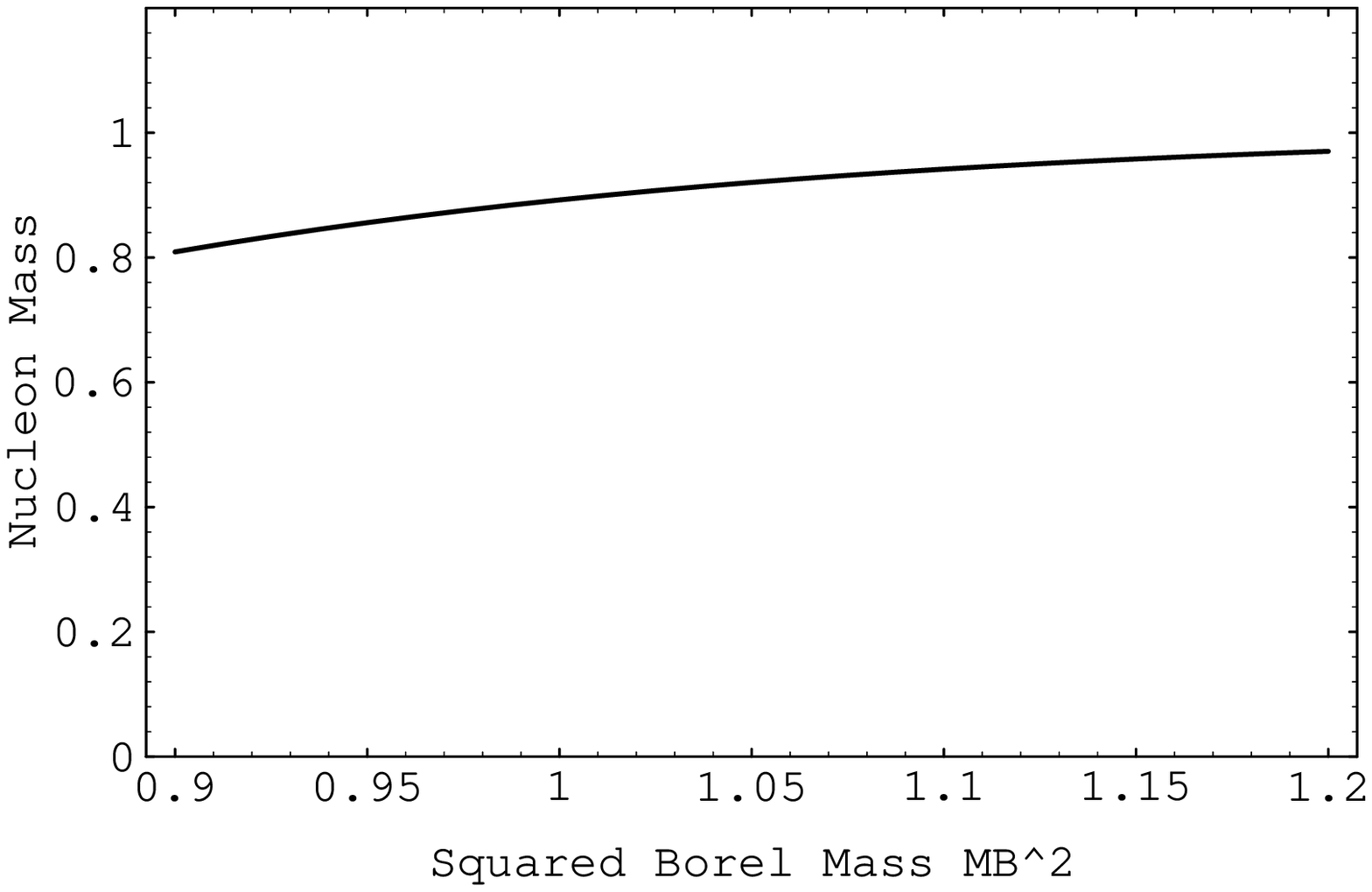,width=6.5in}}

   \end{figure}

   FIG.5

\newpage

   \begin{figure}

     \centerline{\psfig{file=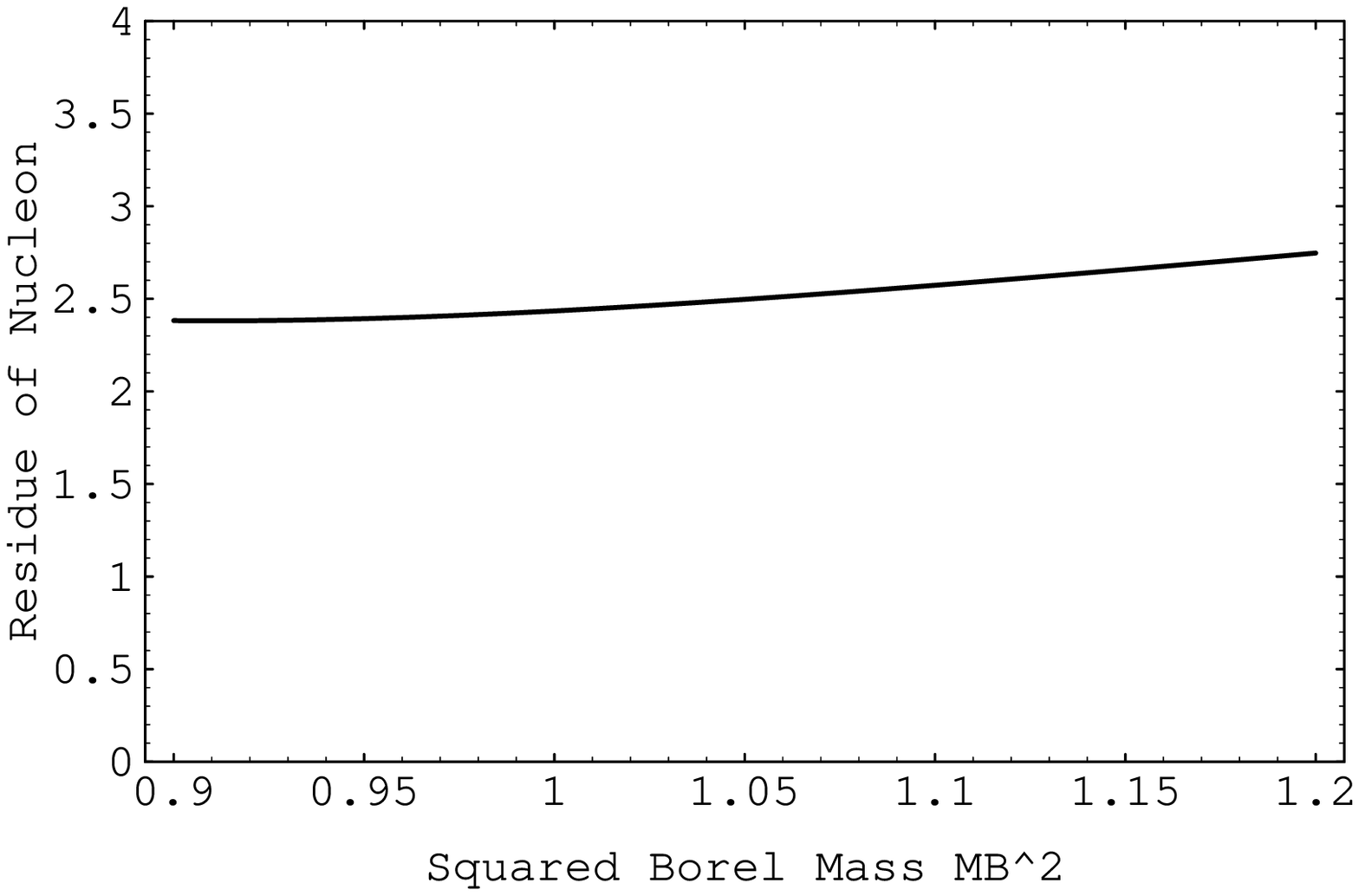,width=6.5in}}

   \end{figure}

   FIG.6

\newpage

   \begin{figure}

     \centerline{\psfig{file=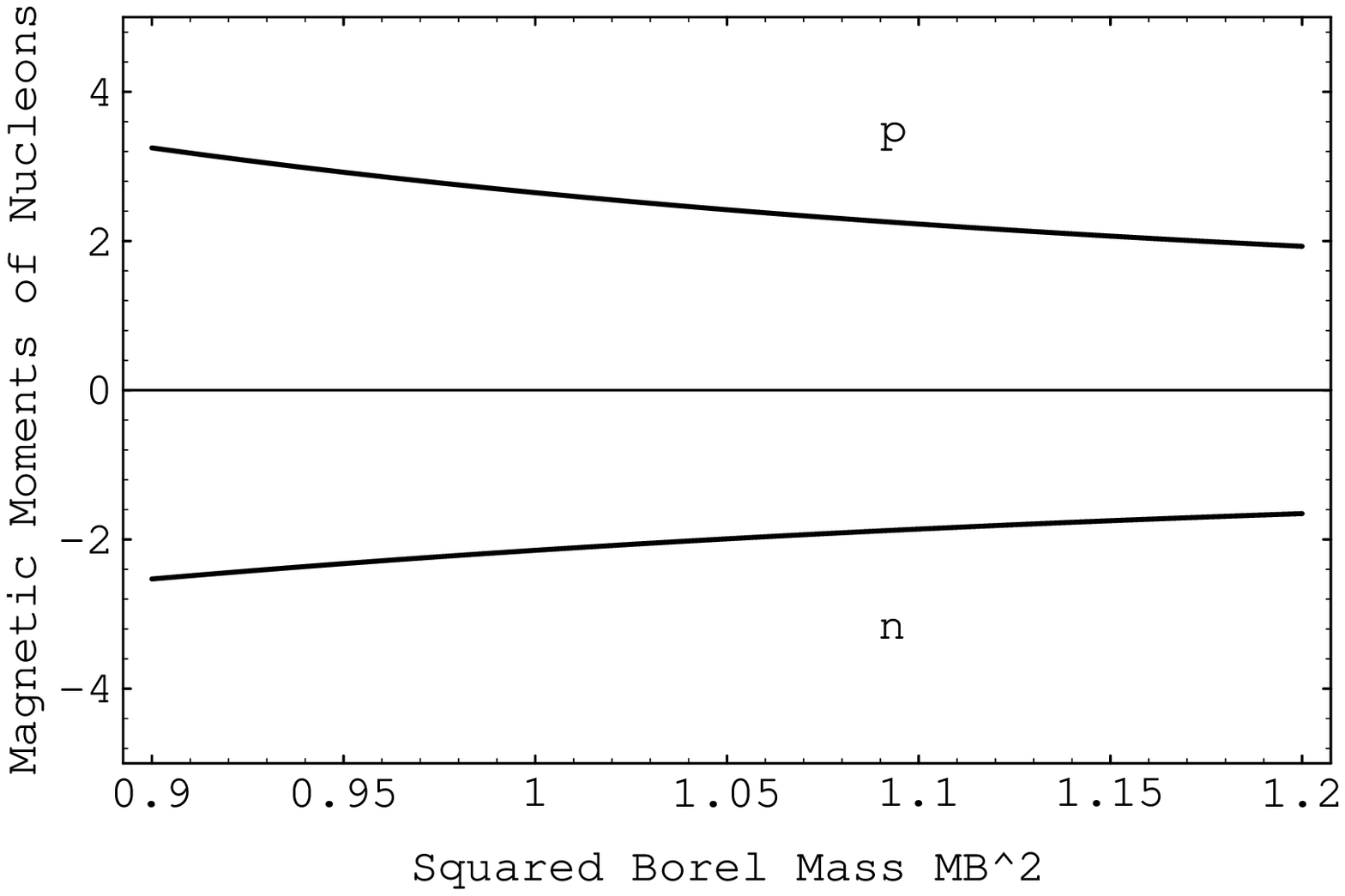,width=6.5in}}
  
   \end{figure}

   FIG.7

 \newpage

   \begin{figure}

     \centerline{\psfig{file=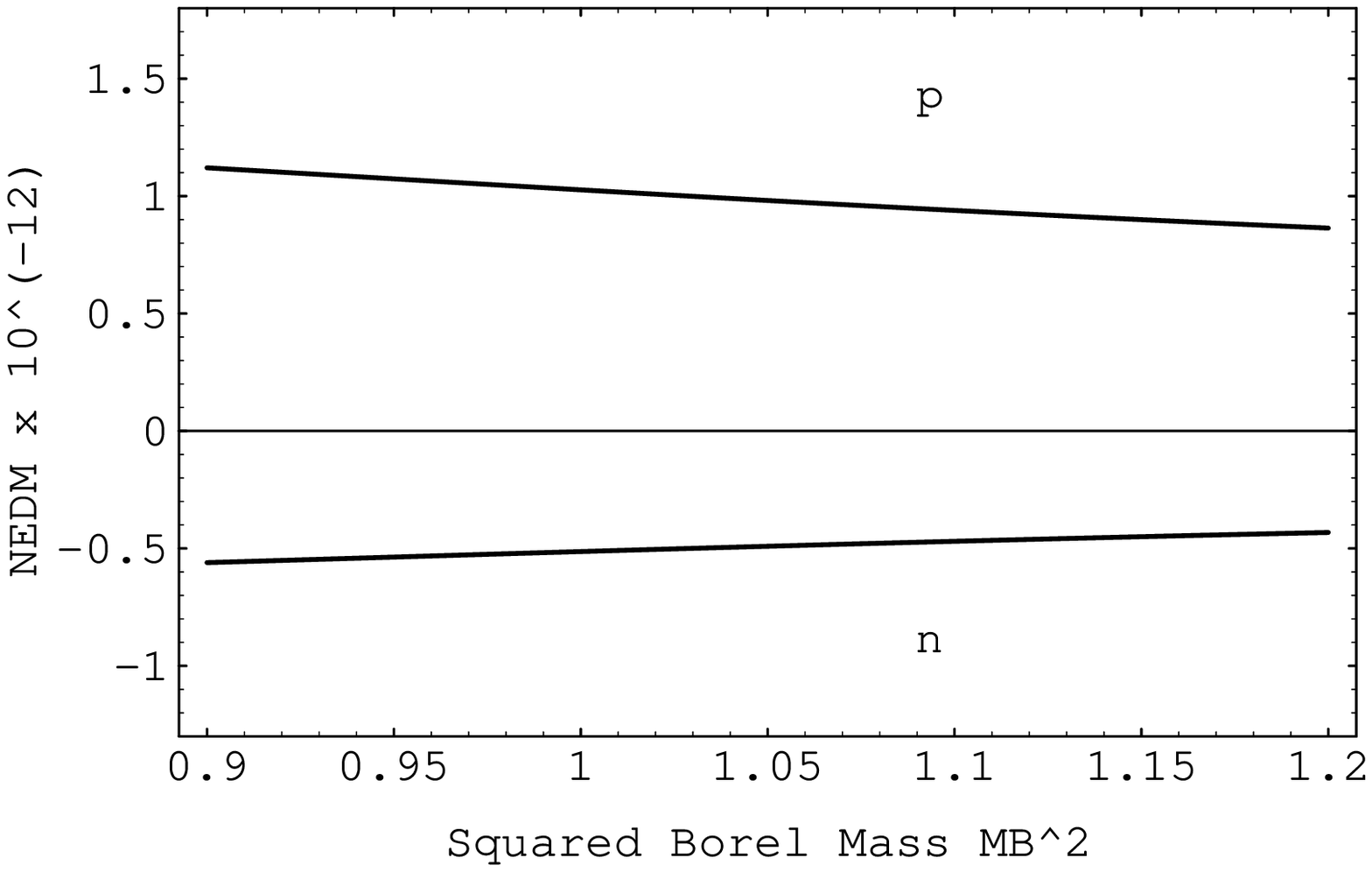,width=6.5in}}

   \end{figure}

   FIG.8

    \newpage

          \[ \begin{array}{|c|c|c|}
             \hline
             \Pi_N^{\mu \nu} & N                  &  N''                \\
             \hline
                   N         & \cond{N|J_\mu|N}   & \cond{N|J_\mu|N''}  \\
                             & Region 1           & Region 2            \\
             \hline
                   N'        & \cond{N'|J_\mu|N}  & \cond{N'|J_\mu|N''} \\
                             & Region 2           & Region 3            \\
             \hline
             \end{array} \]
          TABLE I
 

\begin{references}

\bibitem{NEDM:ramsey}
         N.F. Ramsey, 
         Ann. Rev. Nucl. Part. Sci. {\bf 40}, 1 (1991). 

\bibitem{QCD:theta}
         F.J. Yndurain,
         {\it The theory of quark and gluon interactions,\/}
         (Springer, Berlin; New York, 1993).

         R.K. Bhaduri,
         {\it Models of the Nucleon,\/}
         (Addison-Wesley, Redwood, 1988).

\bibitem{NEDM:exp}
         I.S. Altarev {\it et al.,\/}
         Phys. Lett. B{\bf 276}, 242 (1992).

         K.F. Smith {\it et al.,\/}
         Phys. Lett. B{\bf 234}, 191 (1990).

\bibitem{NEDM:th}
         R.B. Clark and J. Randa, 
         Phys. Rev. D{\bf 12}, 3564 (1975).

         V. Baluni,
         Phys. Rev. D{\bf 19}, 2227 (1979).

         R.J. Crewther, P. Di Vecchia, G. Veneziano, and
         E. Witten, 
         Phys. Lett. B{\bf 88}, 123 (1979), 
         Erratum-ibid. B{\bf 91}, 487 (1980).

         M.M. Musakhanov and Z.Z. Israilov,  
         Phys. Lett. B{\bf 137}, 419 (1984). 

         M.A. Morgan and G.A. Miller, 
         Phys. Lett. B{\bf 179}, 379 (1986).   

         S. Aoki and
         T. Hatsuda,
         Phys. Rev. D{\bf 45}, 2427 (1992).

         L.J. Dixon, A. Langnau, Y. Nir, and B. Warr, 
         Phys. Lett. B{\bf 253}, 459 (1991). 

         H.J. Schnitzer,
         Phys. Lett. B{\bf 253}, 465 (1991). 

         A. Abada {\it et al.,\/}
         J. Galand, A. Le Yaouanc, L. Oliver, O. Pene, J.C. Raynal 
         Phys. Lett. B{\bf 256}, 508 (1991). 


         Hai-Yang Cheng,
         Phys. Rev. D{\bf 44}, 166 (1991). 


         A. Pich and E. de Rafael,
         Nucl. Phys. B{\bf 367}, 313 (1991).

         P. Salomonson, B. Skagerstam, and
         A. Stern, 
         Mod. Phys. Lett. A{\bf 6}, 3647 (1991). 
 
         J.A. McGovern and M.C. Birse, 
         Phys. Rev. D{\bf 45}, 2437 (1992). 

         H.A. Riggs and H.J. Schnitzer,
         Phys. Lett. B{\bf 305}, 252 (1993). 

\bibitem{CP:review}
         G.A. Christos,
         Phys. Rept. {\bf 116}, 251 (1984). 

         Jihn E. Kim,
         Phys. Rept. {\bf 150}, 1 (1987).

         Hai-Yang Cheng, 
         Phys. Rept. {\bf 158}, 1 (1988).

\bibitem{CP:chiral sym1} 
         R.D. Peccei and H.R. Quinn,
         Phys. Rev. Lett. {\bf 38}, 1440 (1977);
       %
         Phys. Rev. D{\bf 16}, 1791 (1977).

\bibitem{CP:chiral sym2}
         E.P. Shabalin,
         Sov. J. Nucl. Phys. {\bf 36}, 575 (1982).
 
\bibitem{CP:chiral sym3}
         M.A. Shifman, A.I. Vainshtein, and V.I. Zakharov, 
         Nucl. Phys. B{\bf 166}, 493 (1980).
 
         S. Aoki, A. Gocksch, A.V. Manohar, and S.R. Sharpe,
         Phys. Rev. Lett. {\bf 65}, 1092 (1990).

         S. Aoki and T. Hatsuda,
         Phys. Rev. D{\bf 45}, 2427 (1992).

\bibitem{CP:chan1}
         Chuan-Tsung Chan, LANL e-Print Archive: hep-ph/9704427.

\bibitem{QSR:nmm}
         We follow the approach in the calculation of the nucleon magnetic
         moments as 
         B.L. Ioffe and A.V. Smilga, 
         Nucl. Phys. B{\bf 232}, 109 (1984).

\bibitem{QCD:ABJ Anomaly}
         S.L. Adler,
         Phys. Rev. {\bf 177}, 2426 (1969).

         J.S. Bell and R. Jackiw,
         Nuovo Cim. {\bf 60}A, 47 (1969).  

\bibitem{QCD:Fuji}
         K. Fujikawa,
         Phys. Rev. Lett. {\bf 42}, 1195 (1979);
         {\it ibid.\/} {\bf 44}, 1733 (1980);
         Phys. Rev. D {\bf 21}, 2848 (1980);
         {\it ibid.\/} {\bf 22}, 1499 (E) (1980).
       %
       
         G.A. Christos,
         Z. Phys. C{\bf 18}, 155 (1983), Erratum-ibid. C{\bf 20}, 186 (1983). 

\bibitem{QSR:current}
         B.L. Ioffe,
         Nucl. Phys. B{\bf 188}, 317 (1981), 
         Erratum-ibid. B{\bf 191}, 591 (1981).

         B.L. Ioffe and A.V. Smilga,
         Nucl. Phys. B{\bf 232}, 109 (1984). 
      
         Y. Chung, H.G. Dosch, M. Kremer, and D. Schall, 
         Nucl. Phys. B{\bf 197}, 55 (1982).

         B.L. Ioffe, 
         Z. Phys. C{\bf 18}, 67 (1983).

         G.A. Christos,
         Z. Phys. C{\bf 29}, 361 (1985). 

\bibitem{QSR:external field}
         B.L. Ioffe, 
         Phys. Atom. Nucl. {\bf 58}, 1408 (1995)
         (e-Print Archive: hep-ph/9501319). 
        
         M. Burkardt,
         D.B. Leinweber,
         and Xue-min Jin, 
         Phys. Lett. B{\bf 385}, 52 (1996). 

\bibitem{math:tensor trick}
         Chuan-Tsung Chan, unpublished note.

\bibitem{QSR:svz}
         M.A. Shifman, A.I. Vainshtein, and V.I. Zakharov,
         Nucl. Phys. B {\bf 147}, 385, (1979);
                       {\bf 147}, 448, (1979);
                       {\bf 147}, 519, (1979).
       %
       %
       %
       %

\bibitem{QSR:chiral sym1}
         R.J. Crewther,
         NATO Advanced Study Inst. on 
         {\it Field Theoretical Methods in Elementary Particle Physics,\/}
         p.529 (1979).

\bibitem{QSR:chiral sym2}
         D.K. Griegel and T.D. Cohen, 
         Phys. Lett. B{\bf 333}, 27 (1994). 

         S-H Lee, S. Choe, 
         T.D. Cohen, and 
         D.K. Griegel, 
         Phys. Lett. B{\bf 348}, 263 (1995). 

\bibitem{QSR:excitation}
         Xue-min Jin, M. Nielsen, and 
         J. Pasupathy, 
         Phys. Rev. D{\bf 51}, 3688 (1995). 

\bibitem{QSR:chan3}
         Chuan-Tsung Chan, work in progress.

\bibitem{QSR:ope}
         A.I. Vainshtein, V.I. Zakharov, V.A. Novikov, and M.A. Shifman,
         Sov. J. Nucl. Phys. {\bf 39}, 77 (1984), Yad. Fiz. {\bf 39}, 124
         (1984). 

         V.A. Novikov, M.A. Shifman, A.I. Vainshtein, and V.I. Zakharov,
         Fortsch. Phys. {\bf 32}, 585 (1985). 

\bibitem{NEDM:chan0}
         Chuan-Tsung Chan, Ph.D. Thesis, University of Washington, 1996.

\bibitem{QSR:DBL}
         D. B. Leinweber
         Ann. Phys. {\bf 254}, 328 (1997). 

    \end{references}
\end{document}